\documentclass[journal]{IEEEtran}
\usepackage{amsmath,amsfonts}
\usepackage{algorithmic}
\usepackage{subfigure}
\usepackage{algorithm}
\usepackage{array}
\usepackage[caption=false,font=normalsize,labelfont=sf,textfont=sf]{subfig}
\usepackage{textcomp}
\usepackage{stfloats}
\usepackage{url}
\usepackage{verbatim}
\usepackage{graphicx}
\usepackage{cite}
\usepackage{bm}
\usepackage{booktabs}
\allowdisplaybreaks[4]

\begin{document}

\title{A Novel OTFS-based Massive Random Access Scheme in Cell-Free Massive MIMO Systems for High-Speed Mobility}

\author{Yanfeng Hu, Dongming Wang,~\IEEEmembership{Member,~IEEE,} Xinjiang Xia,~\IEEEmembership{Member,~IEEE,}  \\Jiamin Li,~\IEEEmembership{Member,~IEEE,} Pengcheng Zhu,~\IEEEmembership{Member,~IEEE,} and Xiaohu You,~\IEEEmembership{Fellow,~IEEE}
\thanks{This work was supported by the National Key R\&D Program of China under Grant 2023YFB2603802, and by the National Natural Science Foundation of China (NSFC) under Grant 62371346. \textit{(Corresponding authors: Dongming Wang.)}}
 \thanks{Y. Hu, D. Wang, J. Fu, J. Li, P. Zhu and X. You are with the National Mobile Communications Research Lab, Southeast University, Nanjing, 210096, P.R.China, and also with the Purple Mountain Laboratories, Nanjing, 210096, P.R.China (email: huyanfeng@seu.edu.cn, wangdm@seu.edu.cn, jiaminli@seu.edu.cn, p.zhu@seu.edu.cn, xhyu@seu.edu.cn)}
 \thanks{X. Xia is with the Purple Mountain Laboratories, Nanjing, 210096, P.R.China (email: xinjiang$\_$xia@aa.seu.edu.cn)}}

\markboth{Journal of \LaTeX\ Class Files,~Vol.~14, No.~8, August~2021}%
{Shell \MakeLowercase{\textit{et al.}}: A Sample Article Using IEEEtran.cls for IEEE Journals}


\maketitle

\begin{abstract}
  In the research of next-generation wireless communication technologies, orthogonal time frequency space (OTFS) modulation is emerging as a promising technique for high-speed mobile environments due to its superior efficiency and robustness in doubly-selective channels. Additionally, the cell-free architecture, which eliminates the issues associated with cell boundaries, offers broader coverage for radio access networks. By combining cell-free network architecture with OTFS modulation, the system may meet the demands of massive random access required by machine-type communication devices in high-speed scenarios. This paper explores a massive random access scheme based on OTFS modulation within a cell-free architecture. A transceiver model for uplink OTFS signals involving multiple access points (APs) is developed, where channel estimation with fractional channel parameters is approximated as a two-dimensional block sparse matrix recovery problem. Building on existing superimposed and embedded preamble schemes, a hybrid preamble strategy intended for massive random access is proposed. This scheme leverages superimposed and embedded preambles to respectively achieve rough and accurate active user equipment (UEs) detection (AUD), as well as precise channel estimation. Moreover, this study introduces a generalized approximate message passing and pattern-coupled sparse Bayesian learning with Laplacian prior (GAMP-PCSBL-La) algorithm, which effectively captures block sparse features after discrete cosine transform (DCT), delivering precise estimation results with reduced computational complexity. Simulation results demonstrate that the proposed scheme is effective and provides superior performance compared to other existing schemes.
\end{abstract}


\section{Introduction}
\IEEEPARstart{T}{he} next-generation wireless communication will delve deeper into more ubiquitous Internet of Things (IoT) scenarios in the coming decades, encompassing broader coverage areas and a significantly larger number of user equipment (UEs) \cite{matthaiou2021road}. Beyond human-type communication devices (HTCDs), there are numerous machine-type communication devices (MTCDs) that need to facilitate data transmission \cite{wu2020massive}. In high-speed massive machine-type communication (mMTC) scenarios, such as high-speed railways, Internet of Vehicles (IoV), unmanned aerial vehicle (UAV) communications, and high-speed integrated sensing and communication (ISAC) \cite{ai20205g}, the great number of MTCDs face constraints such as allocable resources \cite{shahab2020grant}. Due to asynchronous delays and Doppler shifts caused by high-speed mobility, doubly-selective features are often exhibited by the transmission channel. Traditional coordinated access protocols, which involve multiple handshake processes, not only introduce extra delays but also incur significant signaling overhead \cite{chen2018ultra}. Moreover, coordinated orthogonal resources suffer severe orthogonality degradation in doubly-selective channels, thereby reducing system performance \cite{kodheli2020satellite}. Unlike coordinated schemes, grant-free non-orthogonal multiple access (NOMA) allows devices to transmit data without allocated resources. The receiver performs active UE detection (AUD) and channel estimation (CE) based on unique non-orthogonal preamble sequence assigned to each UE \cite{liu2022evolution}. Therefore, grant-free NOMA in uncoordinated access schemes is considered one of the key technologies for mMTC \cite{shahab2020grant}.

Emerging machine-type wireless transmission services impose stringent demands on communication quality in high-mobility scenarios. Orthogonal frequency division multiplexing (OFDM), widely used in 4G and 5G, can eliminate inter-symbol interference caused by time dispersion using a cyclic prefix (CP), but struggles to mitigate frequency dispersion caused by Doppler shifts, leading to inter-carrier interference \cite{jakes1994microwave}. Hadani et al. proposed a novel two-dimensional modulation known as orthogonal time frequency space (OTFS) \cite{hadani2017orthogonal}. Compared to OFDM, OTFS has been proven to significantly improve transmission performance in doubly-selective channels with only a modest increase in system complexity \cite{farhang2017low}. Specifically, OTFS uses a two-dimensional inverse symplectic finite Fourier transform (ISFFT) to map signals from the Doppler-delay (DD) domain to the time-frequency (TF) domain. Unlike OFDM, each signal symbol in OTFS spans the entire TF domain channel, fully exploiting channel diversity and enhancing reliability \cite{wei2021orthogonal}. Additionally, the number of reflectors is considerably smaller than the dimension of transmitted symbols, resulting in sparsity for channel parameters in the DD domain \cite{hadani2017orthogonal}, which simplify the estimation of channel state information (CSI). Given these advantages, OTFS is considered a promising candidate for next-generation broadband communication modulation technology.

In addition, high-mobility communication inevitably requires wide coverage, as UEs may travel significant distances during communication intervals. Cellular network necessitates handovers for high-mobility UEs, increasing the complexity of system processing \cite{wang2022high}. Moreover, boundary effects limit the transmission efficiency for UEs located at the cell edges \cite{zhang2020prospective}. Therefore, a concept named cell-free massive MIMO has been proposed to support denser and wider device coverage, significantly enhancing spectral efficiency and reliability \cite{ngo2017cell}. By deploying numerous access points (APs) across the coverage area, boundary effects are eliminated in cell-free massive system \cite{bjornson2020scalable}. Each AP is equipped with an independent signal processing unit and connected to a central processing unit (CPU) via fronthaul, providing a flexible networking \cite{wang2023full}. Additionally, with UEs being closer to the receiving antennas, signal transmission and processing delays are significantly reduced. Mohammadi et al. theoretically demonstrated that OTFS modulation can achieve superior performance within a cell-free massive MIMO architecture \cite{mohammadi2022cell}. However, numerous challenges still need to be addressed for massive random access in high-mobility scenarios when integrating cell-free massive MIMO.

Recent discussions on OTFS grant-free access schemes for high-mobility mainly focus on low earth orbit (LEO) satellite communication \cite{gao2023grant}. Shen et al. approximated the OTFS channel as a sparse matrix and utilized the low-complexity pattern-coupled sparse Bayesian learning (PCSBL) for AUD and sparse CE \cite{shen2022random}. Zhou et al. designed a novel training sequences aided OTFS (TS-OTFS) transmission protocol for LEO satellite IoT communication and proposed a two-stage AUD and CE method \cite{zhou2022active}. Besides, a high-speed railway IoT active detection method combining tandem spreading multiple access (TSMA) and OTFS was proposed in \cite{ma2021otfs}. By pre-estimating propagation delays, a preamble transmission method was designed in \cite{sinha2020otfs}, allowing UEs to perform pre-compensation. However, existing research fails to address schemes for massive high-mobility MTCD access that incorporate cell-free massive MIMO systems. Moreover, the current CE methods, including the embedded \cite{raviteja2019embedded} and superimposed \cite{mishra2021otfs} pilot schemes, have their limitations: the former incurs high pilot overhead, while the latter has suboptimal estimation performance. Hence, a balanced scheme is required to ensure accurate estimation while reducing overhead.

To address the aforementioned challenges, this paper investigates the AUD and CE schemes for massive random access in cell-free massive MIMO system. Firstly, we establish OTFS uplink signal model in cell-free massive MIMO system. Secondly, a hybrid preamble scheme is designed, where rough AUD is performed by superimposed preambles, and joint accurate AUD and CE are achieved based on embedded preambles. This scheme reduces the overall sparse signal dimension, allowing the system to accommodate more UEs. Finally, we propose a new block sparse matrix recovery algorithm for AUD and CE, named generalized approximate message passing and pattern-coupled sparse Bayesian learning with Laplacian prior (GAMP-PCSBL-La). Simulations demonstrate that this algorithm achieves better estimation performance compared to existing block recovery algorithms. Our contributions are summarized as follows:
\begin{itemize}
  \item We first analyze the massive random access model with OTFS modulation integrating cell-free massive MIMO in this paper. Through mathematical approximation, utilizing uniform planar array (UPA) of antennas and selecting appropriate preamble sequence embedding positions, we model the preamble signals as a two-dimensional (2-D) sparse compressed sensing model in the delay-Doppler-UE-beam domains. Then AUD and CE are transformed into a 2-D block sparse matrix recovery problem.
  \item To address the scale constraints of high-dimensional sparse matrices in compressed sensing{\footnote{The sparse recovery of compressed sensing requires meeting sparsity constraint \cite{candes2006robust}, i.e. $L>C\cdot {{K}_{a}}\log K$, where $L_p$ denotes the length of observed sequences, $K_a$ and $K$ are the number of nonzero and total elements of sparse sequence, respectively. $C$ is a small constant.}} while reducing the overhead of preambles, we propose a hybrid preamble scheme. Rough AUD is performed by superimposed preamble, followed by accurate AUD and CE based on embedded preamble. This approach reduces the sparse channel dimension for each estimation, enabling the system to support massive access for numerous UEs.
  \item A novel GAMP-PCSBL-La algorithm is designed to recover the two-dimensional block sparse channel matrix. GAMP achieves good estimation performance while reducing computational complexity by avoiding matrix inversion \cite{rangan2011generalized}. PCSBL captures the block sparsity of two-dimensional matrix \cite{fang2014pattern}, and the Laplacian prior distribution has been proven to enhance reconstruction ability of sparse signals with discrete cosine transform (DCT) \cite{bellili2019generalized}. By combining these features, GAMP-PCSBL-La achieves excellent channel estimation accuracy with low computational complexity. Our simulation results further validate this conclusion.
\end{itemize}

The remainder of this paper is organized as follows. In Section II, we introduce the system model. Section III discusses the rough AUD, joint accurate AUD and CE strategies based on hybrid preamble scheme. In Section IV, we present the novel block sparse matrix recovery algorithm, GAMP-PCSBL-La. Section V provides numerical simulations and the corresponding analysis. Finally, the conclusion is given in Section VI.

\emph{Notations: }Bold lower letters and bold capital letters denote vectors and matrices, respectively. Normal lower letters and capital letters represent scalar variables and constants, respectively. $\mathbb{C}$ and $\mathbb{R}$ are complex number set and real number set, respectively. $\mathbf{A}_{a:b,c:d}$ represent a sliced matrix for $\mathbf{A}$ with $a$-th row to $b$-th row and $c$-th column to $d$-th column, while $\mathbf{a}_{a:b}$ is a sliced vector for $\mathbf{a}$ with $a$-th element to $b$-th element. Especially, $\mathbf{A}_{a:b,:}$ denotes the submatrix of $\mathbf{A}$ with $a$-th row to $b$-th row. $\mathbb{E}$ and $\mathbb{V}$ mean the expectation and variance, respectively. $\delta(\cdot)$ denotes a Dirac delta function, and $()^H$ is the conjugate transpose of a matrix or vector. $\mathbf{A}[a,b]$ means $(a,b)$-th element of matrix $\mathbf{A}$. $ \otimes $ and $ \odot $ represent Kronecker product and Hadamard product, respectively. Calligraphy letters are used to denote sets. $\left\| \cdot \right\|_F$ is Frobenius norm. $\left\lceil \cdot \right\rceil$ and $\left[ \cdot \right]_{\text{R}}$ respecitvely represent ceiling and rounding operations.
\begin{figure*}[htbp]
  \begin{center}
  {\includegraphics[scale=0.55]{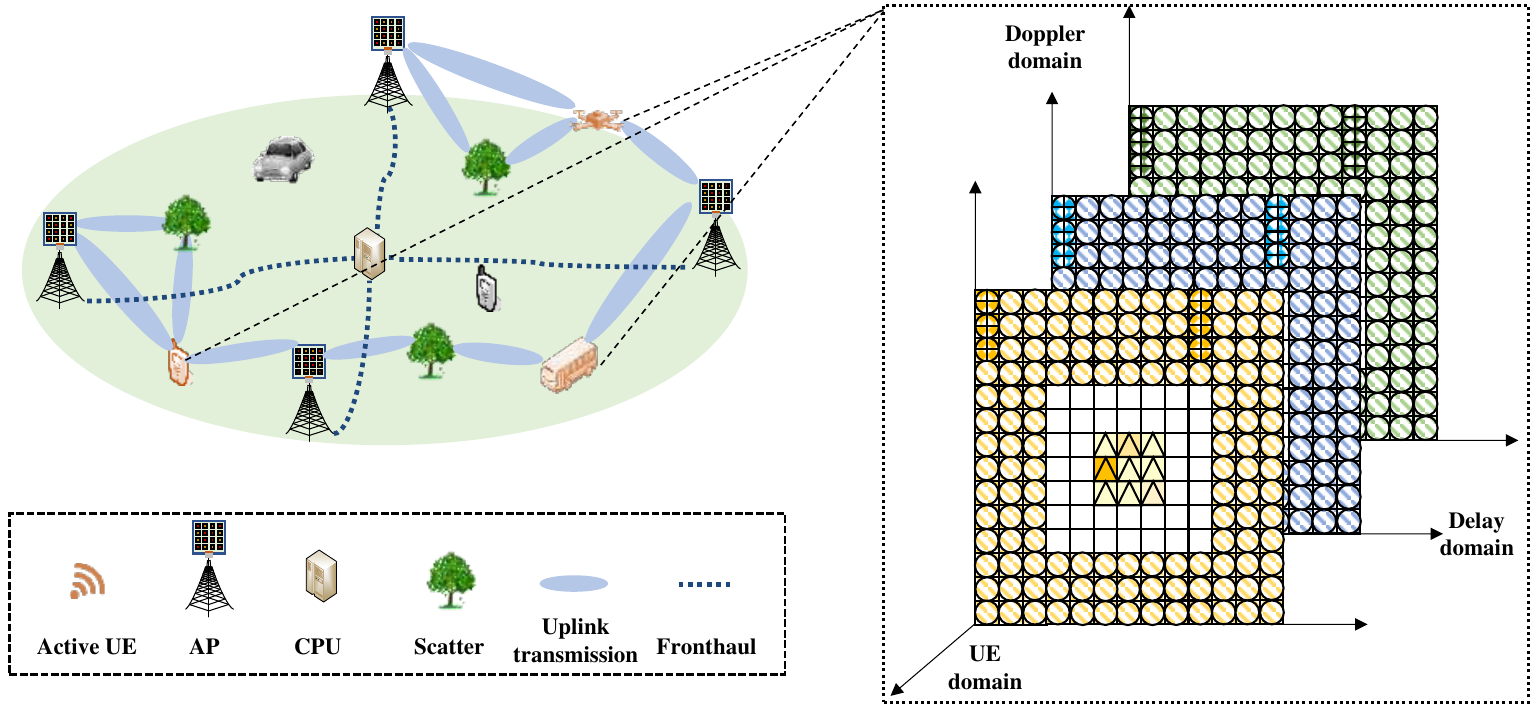}}\\
  \end{center}
  \hfill
  \caption{Massive random access in cell-free massive MIMO system.}
  \label{model}
\end{figure*}
\begin{figure*}[htbp]
  \begin{center}
  {\includegraphics[scale=0.48]{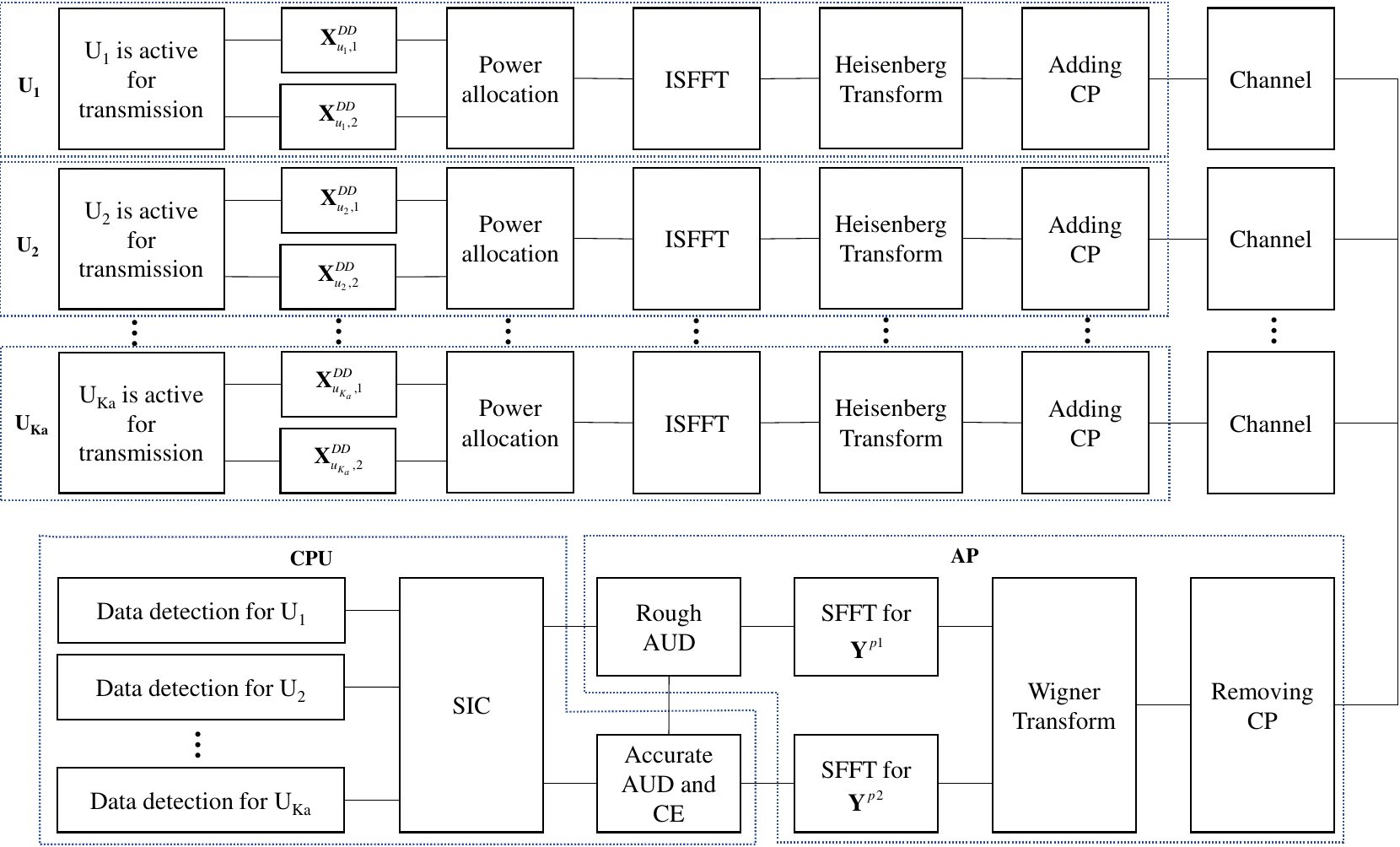}}\\
  \end{center}
  \hfill
  \caption{The system's signal processing flow.}
  \label{proc}
\end{figure*}

\section{System Model}
\subsection{mMTC in Cell-free Massive MIMIO System}
We consider a cell-free massive MIMO system, as shown in Fig. \ref{model}, comprising $B$ APs and $U$ single-antenna UEs, which are randomly distributed over a large area. Assume that each AP is connected to the CPU through fronthaul, allowing lossless data interaction. The UEs move within the area, and only a small portion of UEs transmit uplink data to APs in a specific transmission slot. These UEs are referred to as active UEs, denoted as $\mathcal{K_A}$, while the remaining UEs are silent. The channels of active UEs and APs experience doubly-selective fading. The maximum delay and Doppler shift are set to $\tau_{\max}$ and $\nu_{\max}$, respectively. The signal propagation from an active UE to an AP is characterized by a finite number of paths. The uplink transmission block consists of preamble sequences and data symbols. AUD and CE are performed based on received preamble sequences.
\begin{figure}[htbp]
  \vspace{-0.5em}
  \begin{center}
  {\includegraphics[scale=0.4]{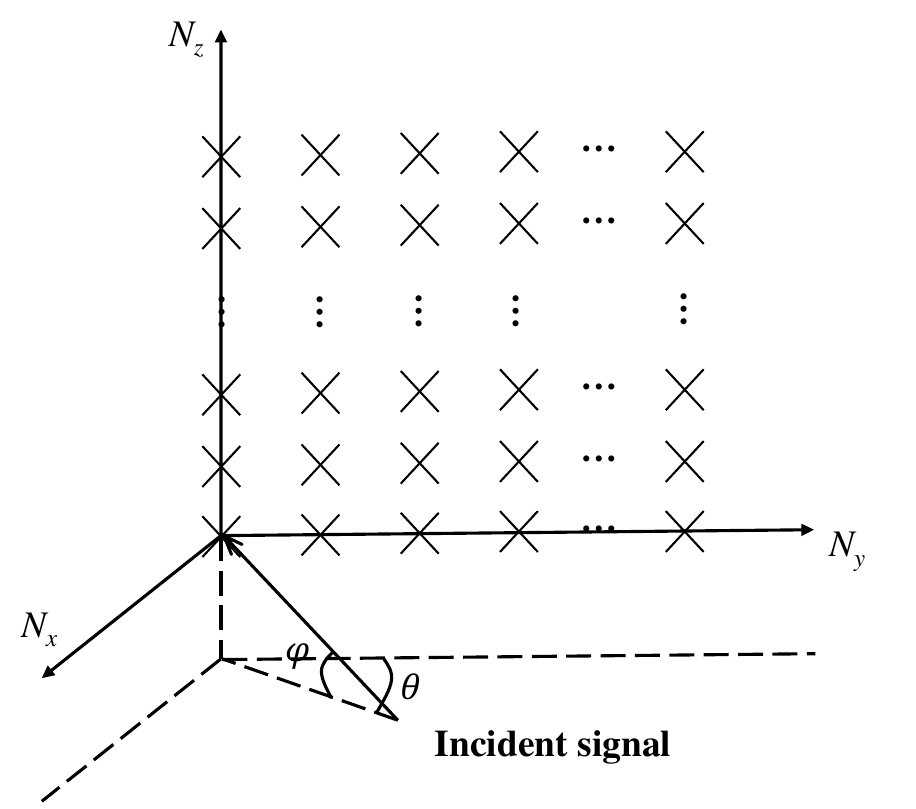}}\\
  \end{center}
  \hfill
  \caption{Structure of UPA equipped at AP.}
  \label{UPA}
  \vspace{-0.8em}
\end{figure}
\vspace{-0.2em}
\subsection{OTFS Modulation and Channel Model}
Consider a typical OTFS transceiver system occupies bandwidth ${M\Delta f}$ and time duration $NT$, where $M$ denotes the number of subcarriers with interval $\Delta f$ and $N$ denotes the number of time intervals $T$. In DD domain, the resolutions of delay and Doppler parameters are $\frac{1}{{M\Delta f}}$ and $\frac{1}{{NT}}$, respectively. For a given active UE $u$, the modulated and power-allocated symbol $\left\{ {X_u^{DD}\left[ {k,l} \right],0 \le k \le N - 1,0 \le l \le M - 1} \right\}$ is assigned to the $(k,l)$-th grid point in $N \times M$ DD grid. Here, $k$ and $l$ represent the indices for Doppler domain and delay domain, respectively. By applying the ISFFT to ${\mathbf{X}}_u^{DD} \in {\mathbb{C}^{N \times M}}$ in DD domain, the $N \times M$ zero-mean symbols are transformed into TF domain:
\begin{equation}\label{Xtfnm}
X_u^{TF}\left[ {n,m} \right] = \frac{1}{{\sqrt {NM} }}\sum\limits_k {} \sum\limits_l {} X_u^{DD}\left[ {k,l} \right]{e^{ - j2\pi \left( {\frac{{ml}}{M} - \frac{{nk}}{N}} \right)}}.
\end{equation}
On this basis, the transmitter applies the Heisenberg transform to convert it into time-domain:
\begin{equation}
{s_u}\left( t \right) = \sum\limits_n {} \sum\limits_m {} \mathbf{X}_u^{TF}\left[ {n,m} \right]{e^{j2\pi m\Delta f(t - nT)}}{g_{tx}}(t - nT),
\end{equation}
where ${g_{tx}}(t)$ represents the rectangular window function in time domain with a duration of $T$. The delay-Doppler channel response model from UE $u$ to the $b$-th AP is defined as:
\begin{small}
\begin{equation}
{h_{u,b}}\left( {\tau ,\nu } \right) = \sum\limits_{i = 1}^P {{h_{u,b,i}}\delta \left( {\tau  - {\tau _{u,b,i}}} \right)\delta \left( {\nu  - {\nu _{u,b,i}}} \right)}.
\end{equation}
\end{small}
Here, ${h_{u,b,i}}$, ${\tau _{u,b,i}}$ and ${\nu _{u,b,i}}$ represent the gain, delay, and Doppler shift, respectively, of the $i$-th path from UE $u$ to the $b$-th AP. $P$ is the number of path. Consider the path loss and shadow fading, we have ${h_{u,b,i}} \sim {\mathcal{C}\mathcal{N}}\left( {0,{\lambda _{u,b}}/{P^2}} \right)$, with ${\lambda _{u,b}}$ representing the large-scale fading coefficient of the channels from UE $u$ to $b$-th AP. The corresponding received time-domain signals for $b$-th AP is presented as:
\begin{small}
\begin{equation}
{r_b}\left( t \right) = \iint {}{h_{u,b}}\left( {\tau ,\nu } \right){s_u}\left( {t - \tau } \right){e^{j2\pi \left( {t - \tau } \right)\nu }}d\tau d\nu  + {n_b}\left( t \right),
\end{equation}
\end{small}
where ${n_b}\left( t \right)$ represents the additive Gaussian noise. Each AP is equipped with a UPA, where $N_z$ and $N_y$ represent the number of antennas in the $z$-direction and $y$-direction, respectively. As shown in Fig. \ref{UPA}, let the elevation angle and azimuth angle of the $i$-th incident signal from the $u$-th active UE to the $b$-th AP are denoted by ${\vartheta _{u,b,i}}$ and ${\varsigma _{u,b,i}}$, respectively. The antenna spacing is specified as half-wavelength, with $\rho _{u,b,i}^y = \cos {\vartheta _{u,b,i}}\cos {\varsigma _{u,b,i}}$ and $\rho _{u,b,i}^z = \sin {\vartheta _{u,b,i}}$. The response of UPA is given by:
\begin{equation}
{\mathbf{a}}^s_{u,b,i} = {{\mathbf{a}}_{{N_y}}}\left( {\rho _{u,b,i}^y} \right) \otimes {{\mathbf{a}}_{{N_z}}}\left( {\rho _{u,b,i}^z} \right),
\end{equation}
where ${{\mathbf{a}}_N}\left( \rho  \right)$ represents the spatial steering vector with dimension $N$, expressed as:
\begin{equation}
{{\mathbf{a}}_N}\left( \rho  \right) = \frac{1}{{\sqrt {{N}} }}{\left[ {1,{\operatorname{e} ^{j\pi \rho }}, \cdots ,{\operatorname{e} ^{j\pi \left( {N - 1} \right)\rho }}} \right]^T}.
\end{equation}
The combine matrix for each AP is denoted as ${\mathbf{W}} = {{\mathbf{D}}_{{N_y}}} \otimes {{\mathbf{D}}_{{N_z}}}$, where ${{\mathbf{D}}_{{N}}}$ is the DFT matrix with dimension $N$. We define beam vector $\mathbf{a}_{u,b,i} = \mathbf{W}^H \mathbf{a}_{u,b,i}^s$, which has only one non-zero block.
The local signal processing unit of AP performs Wigner transform on the time-domain received signal, yielding the received signal presented as
\begin{small}
\begin{align}\label{Ybnm2}
  &{\mathbf{y}_{u,b}^{TF}}\left[ {n,m} \right] = \frac{1}{T}\sum\limits_i {} {h_{u,b,i}}\mathbf{a}_{u,b,i}\sum\limits_{m'} {} {{X}_u^{TF}}\left[ {n,m'} \right]{e^{ - j2\pi m'\Delta f{\tau _{u,b,i}}}}\notag \hfill \\
  &{e^{ - j2\pi {\nu _{u,b,i}}{\tau _{u,b,i}}}}{e^{j2\pi {\nu _{u,b,i}}nT}}\int_{{\tau _{u,b,i}}}^T {} {e^{ - j2\pi \Delta ft(m - m' - \frac{{{\nu _{u,b,i}}}}{{\Delta f}})}}dt \notag \hfill \\
  & +{{X}_u^{TF}}\left[ {n - 1,m'} \right]{e^{ - j2\pi m'\Delta f{\tau _{u,b,i}}}}{e^{j2\pi m'\Delta fT}}{e^{j2\pi {\nu _{u,b,i}}nT}}\notag \hfill \\
  & {e^{ - j2\pi {\nu _{u,b,i}}{\tau _{u,b,i}}}}\int_0^{{\tau _{u,b,i}}} {} {e^{ - j2\pi \Delta ft(m - m' - \frac{{{\nu _{u,b,i}}}}{{\Delta f}})}}dt \notag \hfill \\
  & +{\mathbf{n}_{u,b}}\left[ {n,m} \right] \in \mathbb{C}^{N_zN_y\times1}. 
\end{align}
\end{small}
Usually, $M$ is larger than $N$. We assume that each delay parameter is an integer multiple of the resolution, i.e.,
\begin{small}
\begin{subequations}
\begin{gather}
{\tau _{u,b,i}} = \frac{{{l_{u,b,i}}}}{{M\Delta f}}, \label{tau}\\
{\nu _{u,b,i}} = \frac{{{k_{u,b,i}} + {{\tilde k}_{u,b,i}}}}{{NT}},\label{nu}
\end{gather}
\end{subequations}
\end{small}
where both ${l_{u,b,i}}$ and ${k_{u,b,i}}$ are integers, and ${\tilde k_{u,b,i}}$ is a fraction value between -0.5 and 0.5. Using symplectic finite Fourier transform (SFFT), the received signal in the TF domain is transformed into the DD domain:
\begin{small}
\begin{equation}\label{Yddkl}
  \mathbf{y}_{u,b}^{DD}[k,l] = \frac{1}{{\sqrt {NM} }}\sum\limits_n {} \sum\limits_m {} \mathbf{y}_{u,b}^{TF}\left[ {n,m} \right]{e^{j2\pi \left( {\frac{{ml}}{M} - \frac{{nk}}{N}} \right)}} \in \mathbb{C}^{N_zN_y \times 1}.
\end{equation}
\end{small}
Additionally, we introduce a function $\psi \left( x \right)$ defined as
\begin{small}
\begin{equation}\label{psi}
\psi_N \left( {x,y} \right) = \frac{1}{N}\frac{{1 - {e^{j2\pi x}}}}{{1 - {e^{ - \frac{{j2\pi (y-x)}}{N}}}}}.
\end{equation}
\end{small}
Combining equations (\ref{Xtfnm}), (\ref{Ybnm2}), (\ref{tau}), (\ref{nu}) and (\ref{Yddkl}), we obtain the received signal model in DD domain as $\mathbf{y}_{u,b}^{DD}\left[ {k,l} \right]=\sum_{i=1}^P\mathbf{y}_{u,b,i}^{DD}\left[ {k,l} \right]$, where $\mathbf{y}_{u,b,i}^{DD}\left[ {k,l} \right]$ is shown in equation (\ref{Ydd}), $k'' \in \left[ {{k_{u,b,i}} - \varepsilon ,{k_{u,b,i}} + \varepsilon } \right]$ is defined as the neighborhood of integer Doppler parameters, and $\varepsilon $ is a very small integer. 
\begin{IEEEproof}[Proof of (\ref{Ydd})]
  Please refer to Appendix A in \cite{hu2025novelmassiverandomaccess}.
\end{IEEEproof}
\begin{figure*}[htbp]
  \vspace{-2.5em}
  \centering
  \begin{small}
  \begin{equation}\label{Ydd}
  \mathbf{y}_{u,b,i}^{DD}\left[ {k,l} \right] \approx \left\{ {\begin{array}{*{20}{c}}
    { {h_{u,b,i}}\mathbf{a}_{u,b,i}\sum\limits_{k''} {} X_u^{DD}\left[ {k - k'',l - {l_{u,b,i}}} \right]{e^{ - j2\pi \frac{{k - k''}}{N}}}{e^{j2\pi \frac{{\left( {l - {l_{u,b,i}}} \right)\left( {{k_{u,b,i}} + {{\tilde k}_{u,b,i}}} \right)}}{{NM}}}}\psi_N({k_{u,b,i}} + {{\tilde k}_{u,b,i}},k'')},&{l < {l_{u,b,i}}}, \\
    { {h_{u,b,i}}\mathbf{a}_{u,b,i}\sum\limits_{k''} {} X_u^{DD}\left[ {k - k'',l - {l_{u,b,i}}} \right]{e^{j2\pi \frac{{\left( {l - {l_{u,b,i}}} \right)\left( {{k_{u,b,i}} + {{\tilde k}_{u,b,i}}} \right)}}{{NM}}}}\psi_N({k_{u,b,i}} + {{\tilde k}_{u,b,i}},k'')},&{l \ge {l_{u,b,i}}}.
  \end{array}} \right.
  \end{equation}
\end{small}
\end{figure*}

\vspace{-0.5em}

\section{Hybrid Preamble-based AUD and CE Scheme}
\subsection{Rough Estimation}
We are going to consider the signal model with both delay domain and Doppler domain dimensions are relatively small. Assuming $N' = \alpha N$ and $M' = \beta M$ are both integers, where $0 < \alpha ,\beta  < 1$. The quantization value for the maximum delay ${\tau _{\max }}$ is ${\tilde l'_{\max }} = {\tau _{\max }}M'\Delta f = \beta {\tau _{\max }}M\Delta f \ll 1$ when $\beta$ is particularly small, which implies that any delay parameter $0 < {\tilde l'_{u,b,i}} \le {\tilde l'_{\max }} \ll 1$ is a fractional value. The quantization value for the maximum Doppler shift is ${k'_{\max }} = \left\lceil {{\nu _{\max }}N'T} \right\rceil $. Similar to equation (\ref{Ydd}), we obtain the reception model as:
\begin{small}
\begin{align}\label{Ydd1}
  &\mathbf{y}_{u,b}^{DD}\left[ {k',l'} \right]\mathop  \approx \limits^{(a)} \sum\limits_i {} {h_{u,b,i}}\mathbf{a}_{u,b,i}\sum\limits_{k''} {} {X}_u^{DD}\left[ {k' - k'',l'} \right]  {e^{j2\pi \frac{{l'{{k'}_{u,b,i}}}}{{N'M'}}}} \notag \hfill \\
  &{e^{ - j2\pi \frac{{{{\tilde l'}_{u,b,i}}\left( {{{k'}_{u,b,i}} + {{\tilde k'}_{u,b,i}}} \right)}}{{N'M'}}}}\psi_{N'}({{{k'}_{u,b,i}} + {{\tilde k'}_{u,b,i}}},k'')\psi_{M'}({-\tilde l'}_{u,b,i},0),
\end{align}
\end{small}
where ${\tilde l'_{u,b,i}} = {\tau _{u,b,i}}M'\Delta f$, ${k'_{u,b,i}} = \left[ {{\nu _{u,b,i}}N'T} \right]_{\text{R}}$, ${\tilde k'_{u,b,i}} = {\nu _{u,b,i}}N'T - {k'_{u,b,i}}$ and $k'' \in \left[ {{{k'}_{u,b,i}} - \varepsilon ',{{k'}_{u,b,i}} + \varepsilon '} \right]$ is defined as the neighborhood of ${{k'}_{u,b,i}}$ with $\varepsilon '$ is a small-value integer. Due to $l' < M'$, $ - 0.5 \le {\tilde k'_{u,b,i}} < 0.5$, and especially when $M'$ is very small and $N'$ is larger compared to $M'$, (a) holds approximately true. 
\begin{IEEEproof}[Proof of (\ref{Ydd1})]
  Please refer to Appendix B in \cite{hu2025novelmassiverandomaccess}.
\end{IEEEproof}

For a concise expression, we define function:
\begin{equation}
\begin{aligned}
\mathbf{C}&_ {{\mathbf{x}},k,\varepsilon } = \left[{\text{circ(}}{\mathbf{x}},0{\text{),circ(}}{\mathbf{x}},1{\text{),}} \ldots {\text{,circ(}}{\mathbf{x}},k + \varepsilon {\text{),}} \right.\\
  &\left.\qquad {\text{circ(}}{\mathbf{x}}, - \varepsilon {\text{),circ(}}{\mathbf{x}}, - \varepsilon  + 1{\text{),}} \ldots {\text{,circ(}}{\mathbf{x}}, - 1{\text{)}}\right], \label{Cfunc}
\end{aligned}
\end{equation}
where ${\text{circ(}}{\mathbf{x}},i{\text{)}}$ represents the vector obtained by circularly shifting vector $\mathbf{x}$ by $i$ positions. Based on the above definitions, we transform equation (\ref{Ydd1}) into matrix form:
\begin{small}
\begin{equation}\label{ybp1}
\begin{gathered}
  {\mathbf{Y}}_{u,b}^{p1} \approx \left( {{\mathbf{X}}_u^{p1} \odot {\mathbf{\Phi '}}} \right){\mathbf{H}}_{u,b}^{DD1} + {\mathbf{N}}_{u,b}^{DD1} \hfill \\
   = {\mathbf{A}}_u^{p1}{\mathbf{H}}_{u,b}^{DD1} + {\mathbf{N}}_{u,b}^{DD1} \in {\mathbb{C}^{N'M' \times N_yN_z}}, \hfill \\
\end{gathered}
\end{equation}
\end{small}
where ${\mathbf{Y}}_b^{p1} = \left [ \mathbf{y}_b^{DD}\left[ {0,0} \right],...,\mathbf{y}_b^{DD}\left[ {N'-1,M'-1} \right]\right]^T$, and
\begin{small}
\begin{subequations}
\begin{gather}
{\mathbf{X}}_u^{p1} = \left[ {\begin{array}{*{20}{c}}
  {\mathbf{C}_ {\left({\mathbf{X}}_u^{DD} \right)_ {:,1},{{k'}_{\max }},\varepsilon '} } \\
  {\mathbf{C}_ {\left({\mathbf{X}}_u^{DD} \right)_ {:,2},{{k'}_{\max }},\varepsilon '} } \\
   \vdots  \\
  {\mathbf{C}_ {\left({\mathbf{X}}_u^{DD} \right)_ {:,M'},{{k'}_{\max }},\varepsilon '} }
  \end{array}} \right], \\
{\mathbf{\Phi '}} = {\left( {{\mathbf{F}}_{M',N'}} \right)_{:,{{\mathbf{p}}_{r1}}}} \otimes {{\mathbf{1}}^{N' \times 1}}.
\end{gather}
\end{subequations}
\end{small}
Here, ${{\mathbf{F}}_{M',N'}}$ is a matrix, where each element is given by $F_{M',N'}[k,l] = e^{j2\pi \frac{kl}{M'N'}}$. ${{\mathbf{p}}_{r1}}$ is the index vector that satisfies ${{\mathbf{p}}_{r1}} = [1:k'_{\max}+\varepsilon '+1] \cup [N'-\varepsilon '+1:N']$.
${\mathbf{H}}_{u,b}^{DD1} \in {\mathbb{C}^{({{k'}_{\max }} + 2\varepsilon ' + 1) \times 1}}$ is expressed as
\begin{small}
\begin{equation}
{\mathbf{H}}_{u,b}^{DD1} = \sum\limits_i {{\mathbf{H}}_{u,b,i}^{DD1}}.
\end{equation}
\end{small}
\begin{figure*}[htbp]
  \vspace{-2em}
  \centering
  \begin{small}
  \begin{equation}\label{hubidd}
  \left(\mathbf{H}_{u,b,i}^{DD1}\right)_{t,:}  = \left\{ {\begin{array}{*{20}{c}}
    {{\hat h'_{u,b,i}}\mathbf{a}_{u,b,i}^T}\psi_{N'}({ {{\tilde k'}_{u,b,i}}},t'),&{\left\{ {\begin{array}{*{20}{c}}
    {{\text{if }}{{k'}_{u,b,i}} + t' < 0{\text{,  }}t = ({{k'}_{\max }} + 2\varepsilon ' + 1) + {{k'}_{u,b,i}} + t' + 1}, \\
    {{\text{if }}{{k'}_i} + t' \ge 0{\text{,  }}t = {{k'}_{u,b,i}} + t' + 1},
  \end{array}} \right.} \\
    \mathbf{0}^{1\times N_yN_z },&{{\text{otherwise}}}.
  \end{array}} \right.
  \end{equation}
  \end{small}
  \hrulefill
  \vspace{-1.8em}
\end{figure*}
${{\mathbf{H}}_{u,b,i}^{DD1}}$ is presented in equation (\ref{hubidd}), where $ - \varepsilon ' \le t' \le \varepsilon '$ and $ {\hat h'_{u,b,i}} = { h_{u,b,i}}\psi_{M'}({-\tilde l'}_{u,b,i},0){e^{ - j2\pi \frac{{{{\tilde l'}_{u,b,i}}\left( {{{k'}_{u,b,i}} + {{\tilde k'}_{u,b,i}}} \right)}}{{N'M'}}}}$. Combining equations (\ref{ybp1}) and (\ref{hubidd}), ${\mathbf{A}}_u^{p1} = {{\mathbf{X}}_u^{p1} \odot {\mathbf{\Phi '}}}$ is regarded as the known measurement matrix at the AP, ${\mathbf{Y}}_b^{p1}$ as the observed matrix and ${\mathbf{H}}_{u,b}^{DD1}$ as an unknown 2-D block sparse matrix. In a multi-user scenario, the dimension of the sparse matrix expands, allowing us to utilize this model to detect the indices of non-zero entries in the sparse matrix and thereby identify potential active UEs. Given the coarse approximations made during rough AUD, especially under conditions where $M'$ is notably small, accurate estimation of channel parameters becomes challenging. Therefore, it necessitates further refinement based on initial rough detection for accurate AUD and CE. Detailed elaboration on this can be found in following subsections.
\vspace{-1.3em}

\subsection{Accurate Estimation}
\begin{figure}[htbp]
  \begin{center}
  {\includegraphics[scale=0.3]{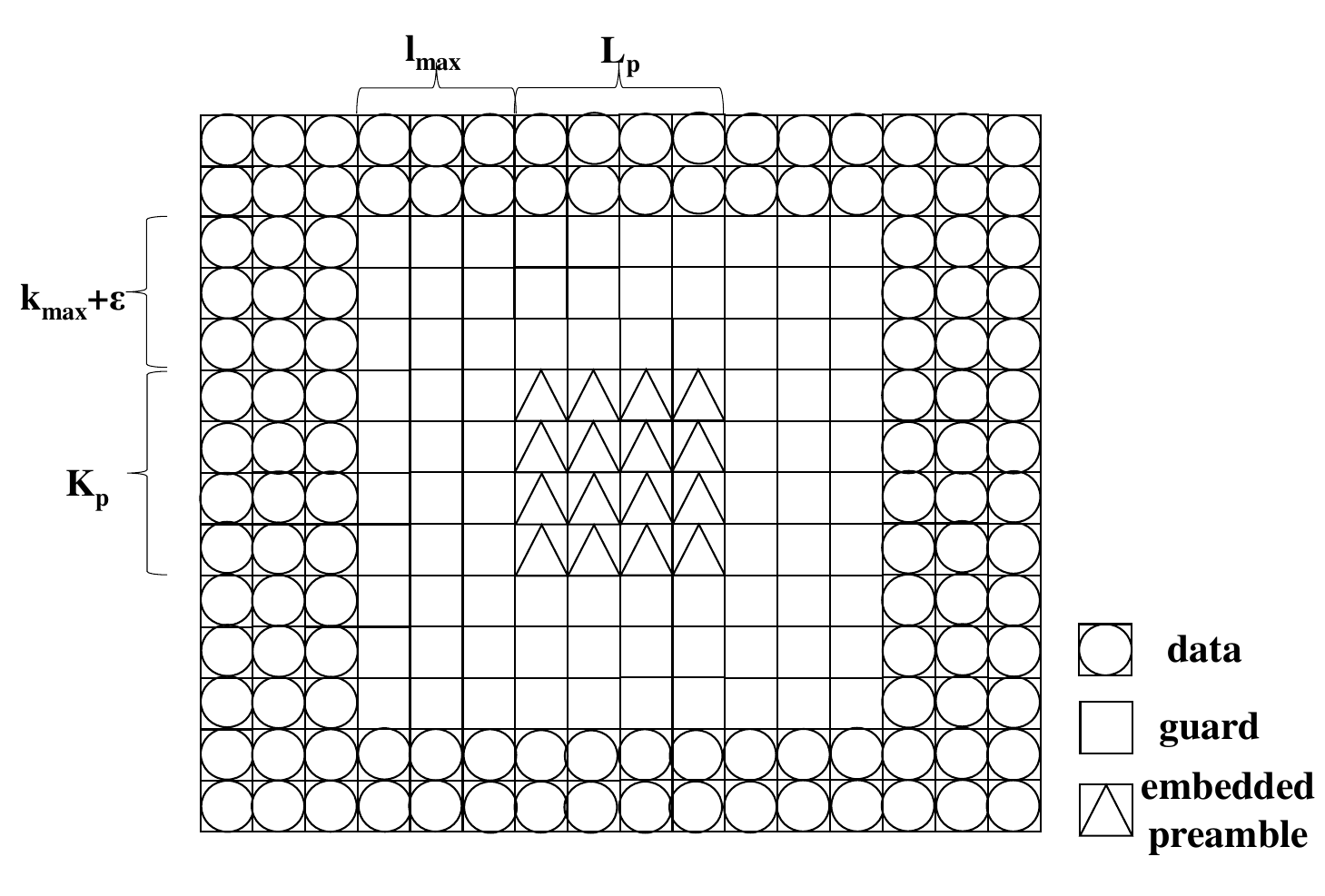}}\\
  \end{center}
  \hfill
  \caption{Symbols arrangement for data, guard and embedded preamble.}
  \label{frame}
  \vspace{-1em}
\end{figure}

Given a sufficiently large dimensions of transmission block, the embedded preamble can be adopted for joint accurate AUD and CE. Let ${k_{\max }} = \left\lfloor {{\nu _{\max }}NT} \right\rfloor $ and ${l_{\max }} = {\tau _{\max }}M\Delta f$. It can be observed from (\ref{Ydd}) that the $(k, l)$-th DD domain received symbol is affected by the transmitted symbols with range of $[k - {k_{\max }} - \varepsilon :k + \varepsilon ,l - {l_{\max }}:l]$. Therefore, to avoid interference between preamble and data, a guard interval needs to be estabilshed, where symbols within this interval are set to zero, as illustrated in Fig. \ref{frame}.

Assuming the starting coordinates of the preamble are $(k_p, l_p)$, with the dimension of $L_p$ on the delay axis and $K_p$ on the Doppler axis. If we set $l_p-l_{\max} \ge l_{\max}$ and $l_p+L_p<M$, for $l \in \left[ {{l_p} - {l_{\max }},{l_p} + {L_p}} \right]$, only the case when $l \ge {l_{u,b,i}}$ (since $l \ge {l_{\max }} \ge {l_{u,b,i}}$) in (\ref{Ydd}) are considerable. We denote $M_p = L_p+l_{\max}$, $N_p = K_p+k_{\max}$, and let ${{\mathbf{X}}_{u,p}} = ({{\mathbf{X}}_u})_{{k_p}:{k_p} + {N_p} + \varepsilon-1 ,{l_p}:{l_p} + M_p-1}$, ${\mathbf{Y}}_{u,b}^{p2} = ({{\mathbf{Y}}_{u,b}})_{\text{vec}({k_p}:{k_p} + {N_p} -1 ,{l_p}:{l_p} + M_p-1),:}$, where $\mathbf{X}_u \in \mathbb{C}^{N \times M}$ and $\mathbf{Y}_{u,b} \in \mathbb{C}^{NM\times N_zN_y}$ are DD domain transmitted symbols of $u$-th UE and received symbols of $b$-th AP, respectively. $\text{vec}( {k_p}:{k_p} + {N_p} -1 ,{l_p}:{l_p} + M_p-1)$ denotes the vectorized matrix indices $[{k_p}:{k_p} + {N_p} -1 ,{l_p}:{l_p} + M_p-1]$.
Similar to (\ref{ybp1}), the received embedded preamble signal is expressed as:
\begin{equation}\label{ybp2}
\begin{gathered}
  {\mathbf{Y}}_{u,b}^{p2} \approx \left( {{\mathbf{X}}_u^{p2} \odot {\mathbf{\Phi }}} \right){\mathbf{H}}_{u,b}^{DD2} + {\mathbf{N}}_{u,b}^{DD2} \hfill \\
   = {\mathbf{A}}_u^{p2}{\mathbf{H}}_{u,b}^{DD2} + {\mathbf{N}}_{u,b}^{DD2} \in {\mathbb{C}^{{N_p}{M_p} \times N_zN_y}}, \hfill \\
\end{gathered}
\end{equation}
where ${\mathbf{X}}_u^{p2}$ is described in equation (\ref{Xup2}), and
\begin{figure*}[ht]
  \vspace{-1.9em}
  \centering
  \begin{small}
    \begin{equation}\label{Xup2}
      \begin{gathered}
        {\mathbf{X}}_u^{p2} = \left[ {\begin{array}{*{20}{c}}
        {{\mathbf{C}^{c}}_ {\left({{\mathbf{X}}_{u,p}} \right)_{:,1},{k_{\max }},\varepsilon } } \\
        {{\mathbf{C}^{c}}_ {\left({{\mathbf{X}}_{u,p}} \right)_{:,2},{k_{\max }},\varepsilon } } \\
         \vdots  \\
         {{\mathbf{C}^{c}}_ {\left({{\mathbf{X}}_{u,p}} \right)_{:,{L_p} + {l_{\max }}},{k_{\max }},\varepsilon } }
      \end{array}\begin{array}{*{20}{c}}
        {{\mathbf{C}^{c}}_ {\left({{\mathbf{X}}_{u,p}} \right)_{:,{L_p} + {l_{\max }}},{k_{\max }},\varepsilon } } \\
        {{\mathbf{C}^{c}}_ {\left({{\mathbf{X}}_{u,p}} \right)_{:,1},{k_{\max }},\varepsilon } } \\
         \vdots  \\
        {{\mathbf{C}^{c}}_ {\left({{\mathbf{X}}_{u,p}} \right)_{:,{L_p} + {l_{\max }}-1},{k_{\max }},\varepsilon } }
      \end{array}\begin{array}{*{20}{c}}
        {} \\
        {} \\
         \cdots  \\
        {}
      \end{array}\begin{array}{*{20}{c}}
        {{\mathbf{C}^{c}}_ {\left({{\mathbf{X}}_{u,p}} \right)_{:,{L_p} + 1},{k_{\max }},\varepsilon } } \\
        {{\mathbf{C}^{c}}_ {\left({{\mathbf{X}}_{u,p}} \right)_{:,{L_p} + 2},{k_{\max }},\varepsilon } } \\
         \vdots  \\
        {{\mathbf{C}^{c}}_ {\left({{\mathbf{X}}_{u,p}} \right)_{:,{L_p}},{k_{\max }},\varepsilon } }
      \end{array}} \right] \hfill \\
      \end{gathered}
    \end{equation}
    \end{small}
    \vspace{-1.8em}
\end{figure*}
\begin{small}
\begin{subequations}
\begin{gather}
{\mathbf{C}^{c}}_{{\mathbf{x}},k,\varepsilon }  = {\left( {\mathbf{C}_ {{\mathbf{x}},k,\varepsilon } } \right)_{1:\dim ({\mathbf{x}}) - \varepsilon ,:}}, \\
{\mathbf{\Phi }} = {{\mathbf{1}}^{1 \times \left( {{l_{\max }} + 1} \right)}}\otimes{\left( {{\mathbf{F}}_{M,N}^{}} \right)_{{l_p}:{l_p} + {M_p} - 1,{{\mathbf{p}}_{r2}}}} \otimes {{\mathbf{1}}^{{N_p} \times 1}}.
\end{gather}
\end{subequations}
\end{small}
${{\mathbf{p}}_{r2}}$ is the index vector that satisfies ${{\mathbf{p}}_{r2}} = [1:k_{\max}+\varepsilon+1] \cup [N_p-\varepsilon+1:N_p]$. ${\mathbf{H}}_{u,b}^{DD2} \in {\mathbb{C}^{({k_{\max }} + 2\varepsilon  + 1)({l_{\max }} + 1) \times N_zN_y}}$ is expressed as
\begin{small}
\begin{equation}
{\mathbf{H}}_{u,b}^{DD2} = \sum\limits_i {{\mathbf{H}}_{u,b,i}^{DD2}}.
\end{equation}
\end{small}
\begin{figure*}[ht]
  \centering
  \begin{small}
    \begin{equation}\label{hubidd2}
      \left(\mathbf{H}_{u,b,i}^{DD2}  \right)_{t,:} = \left\{ {\begin{array}{*{20}{c}}
  {{\hat h_{u,b,i}}\mathbf{a}_{u,b,i}^T\psi_{N}({ {{\tilde k}_{u,b,i}}},t'),}&{\left\{ {\begin{array}{*{20}{c}}
  {{\text{if }}{k_{u,b,i}} + t' < 0{\text{,  }}t = ({l_i} + 1)({k_{\max }} + 2\varepsilon  + 1) + {k_{u,b,i}} + t + 1}, \\
  {{\text{if }}{k_{u,b,i}} + t' \ge 0{\text{,  }}t = {l_i}({k_{\max }} + 2\varepsilon  + 1) + {k_{u,b,i}} + t' + 1},
\end{array}} \right.} \\
  \mathbf{0}^{1\times N_yN_z},&{{\text{otherwise}}}.
\end{array}} \right.
    \end{equation}
    \end{small}
  \hrulefill
  \vspace{-1.8em}
\end{figure*}
${{\mathbf{H}}_{u,b,i}^{DD2}}$ is presented as in equation (\ref{hubidd2}), where $ - \varepsilon  \le t' \le \varepsilon $ and $ {\hat h_{u,b,i}} = { h_{u,b,i}}{e^{ - j2\pi \frac{{{{ l}_{u,b,i}}\left( {{{k}_{u,b,i}} + {{\tilde k}_{u,b,i}}} \right)}}{{NM}}}}$. According to equation (\ref{ybp2}), ${\mathbf{A}}_u^{p2} = {{\mathbf{X}}_u^{p2} \odot {\mathbf{\Phi }}}$ can be considered as the known measurement matrix at the AP, ${\mathbf{Y}}_{u,b}^{p2}$ as the observed matrix and ${\mathbf{H}}_{u,b}^{DD2}$ as an unknown 2-D block sparse matrix. Since in accurate estimation, $M > M'$ and $N > N'$, which make a higher resolution in delay and Doppler, resulting in more precise quantization. However, compared to rough estimation, the dimension of the sparse vector for accurate estimation is larger, making it more difficult to recover the sparse vector in multi-UE scenarios. Therefore, a hybrid preamble scheme is designed to achieve precise detection and estimation with lower overhead and complexity.

\vspace{-1em}
\subsection{Hybrid Preamble for Multi-UE Joint Active Detection and Channel Estimation}\label{secC}
\begin{figure}[htbp]
  \vspace{-1em}
  \begin{center}
  {\includegraphics[scale=0.255]{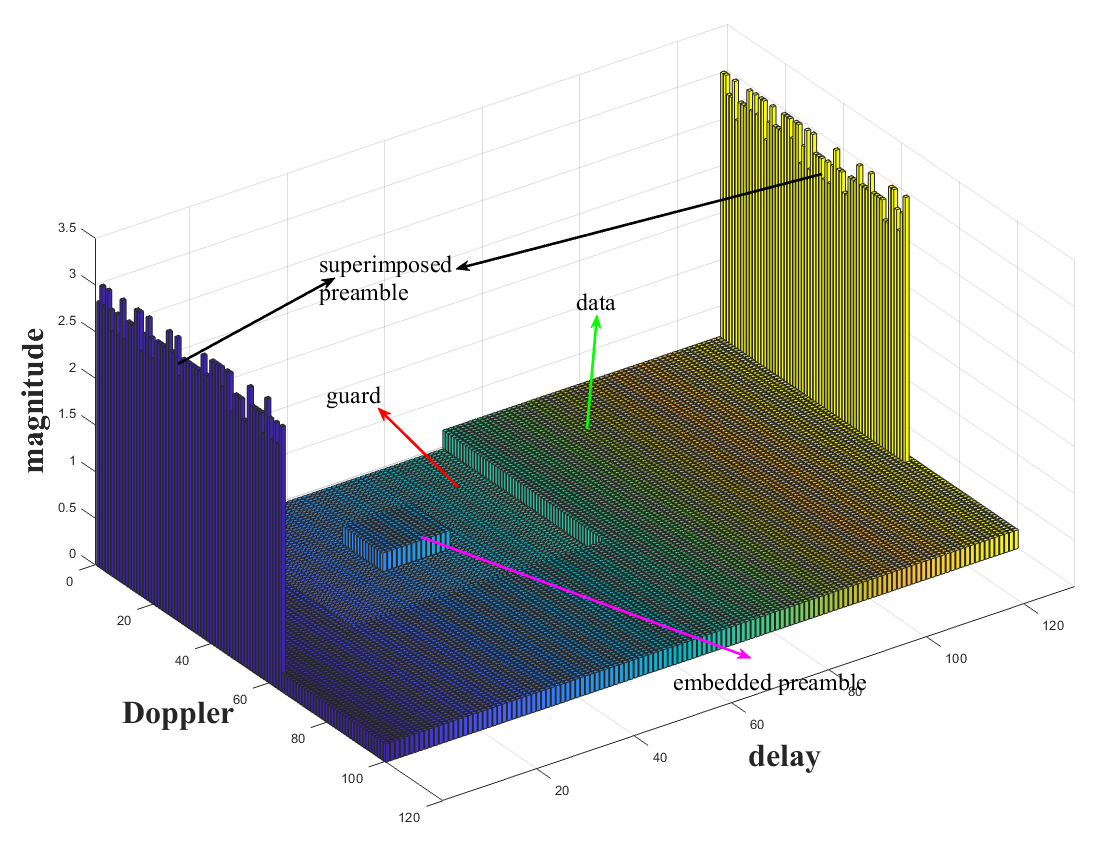}}\\
  \end{center}
  \hfill
  \caption{The hybrid preamble transmission block structure.}
  \label{supStructure}
  \vspace{-1em}
\end{figure}

In the $N \times M$ DD domain, we superimpose the superimposed preamble, denoted as preamble1 ${\mathbf{X}}_{u,p1}^{DD}$, and the block symbols ${\mathbf{X}}_{u,2}^{DD}$, which includes the embedded preamble denoted as preamble2 ${\mathbf{X}}_{u,p2}$, and data symbols ${\mathbf{X}}_{u,d}$. Different power levels are allocated to ${\mathbf{X}}_{u,p1}^{DD}$ and ${\mathbf{X}}_{u,2}^{DD}$, ensuring a significant difference in energy domain between these two types of signals. The superimposed result forms transmission block, structured as shown in the Fig. \ref{supStructure}.
Then we have
\begin{small}
\begin{equation}
\begin{gathered}
  {\mathbf{X}}_u^{DD} = {\mathbf{X}}_{u,1}^{DD} + {\mathbf{X}}_{u,2}^{DD}, \hfill \\
  {{X}}_{u,1}^{DD}[k,l] = \left\{ {\begin{array}{*{20}{c}}
  {{{X}}_{u,p1}^{DD}[k',l']}&{k = k',l = \frac{{l'}}{\beta }}, \\
  0&{{\text{otherwise}}},
\end{array}} \right. \hfill \\
  {{X}}_{u,2}^{DD}[k,l] = \left\{ {\begin{array}{*{20}{c}}
  {{{X}}_{u,p2}^{DD}[k',l']}&{k = k' + {k_p},l = l' + {l_p}}, \\
  {{{X}}_{u,d}^{DD}[k,l]}&{\left( {k,l} \right){\text{ not in }}{\mathcal{P}\mathcal{G}}{\text{ area}}}, \\
  0&{{\text{otherwise}}},
\end{array}} \right. \hfill \\
\end{gathered}
\end{equation}
\end{small}
where $\mathcal{PG}$ area represents the grids designated for placing preamble2 and the guard intervals. We set the preamble1 is placed at intervals of $\frac{1}{\beta }$ along the delay axis while being placed continuously along the Doppler axis. Since the delay dimension $M' = \beta M$ of preamble1 is assumed to be very small, and the Doppler dimension satisfies $N' = \alpha N \le \frac{N}{2}$, there is sufficient space within DD dimension to place preamble2 and guard interval. This arrangement ensures that the received signal of preamble1 does not interfere with preamble2.

Building on this,the received signal in the TF domain can be expressed as:
\begin{small}
\begin{equation}
\begin{gathered}
  {\mathbf{Y}^{TF}_b} = {\mathbf{Y}^{TF}_{b1}} + {\mathbf{Y}^{TF}_{b2}} + \mathbf{N}^{TF}_b  \\
    =  {\mathbf{Y}^{TF}_{b1}} + \mathbf{\tilde Z}^{TF}_b, \\
\end{gathered}
\end{equation}
\end{small}
where ${\mathbf{Y}^{TF}_{b1}}$ and ${\mathbf{Y}^{TF}_{b2}}$ represent $b$-th AP's received ${\mathbf{X}}_{1}^{DD}$ and ${\mathbf{X}}_{2}^{DD}$ signals in TF domain, respectively. $\mathbf{\tilde Z}^{TF}_b = {\mathbf{Y}^{TF}_{b2}} + \mathbf{N}^{TF}_b$ is treated as noise. Assuming that applying ISFFT to ${\mathbf{X}}_{u,p1}^{DD} \in {\mathbb{C}^{N' \times M'}}$ results in a TF domain signal ${\mathbf{X}}_{u,p1}^{TF} \in {\mathbb{C}^{N' \times M'}}$, and applying ISFFT to ${\mathbf{X}}_{u,1}^{DD} \in {\mathbb{C}^{N \times M}}$ to obtain a TF domain signal ${\mathbf{X}}_{u,1}^{TF} \in {\mathbb{C}^{N \times M}}$, both signals pass through the same channel to arrive at the $b$-th AP. After performing the Wigner transform, the TF domain received signals are ${\mathbf{Y}}_{u,b,p1}^{TF} \in {\mathbb{C}^{N' M'\times N_zN_y}}$ and ${\mathbf{Y}}_{u,b,1}^{TF} \in {\mathbb{C}^{N M\times N_zN_y}}$ respectively. Based on equations (\ref{Xtfnm}), (\ref{Yddkl}), and (\ref{hubidd2}), we can derive:
\begin{small}
\begin{equation}
{{X}}_{u,p1}^{TF}[n',m'] = \frac{1}{{\sqrt {\alpha \beta } }}{{X}}_{u,1}^{TF}[\frac{{n'}}{\alpha },m'],
\end{equation}
\end{small}
\begin{small}
\begin{equation}\label{Yub1}
\begin{gathered}
  {\mathbf{y}}_{u,b,1}^{TF}[\frac{{n'}}{\alpha },m'] = \sqrt {\alpha \beta } \frac{1}{T}\sum\limits_i {} {h_{u,b,i}\mathbf{a}_{u,b,i}}\sum\limits_{m''} {} {{X}}_{u,p1}^{TF}[n',m''] \hfill \\
  {e^{ - j2\pi m''\Delta f{\tau _{u,b,i}}}}{e^{ - j2\pi {\nu _{u,b,i}}{\tau _{u,b,i}}}}{e^{j2\pi {\nu _{u,b,i}}\frac{{n'}}{\alpha }T}} \hfill \\
  \int_{{\tau _{u,b,i}}}^T {} {e^{ - j2\pi \Delta ft(m' - m'' - \frac{{{\nu _{u,b,i}}}}{{\Delta f}})}}dt + \mathbf{n}_{u,b,1}^{TF}[\frac{{n'}}{\alpha },m'], \hfill \\
\end{gathered}
\end{equation}
\end{small}
\begin{small}
\begin{equation}\label{Yubp1}
\begin{gathered}
  {\mathbf{y}}_{u,b,p1}^{TF}[n',m'] = \frac{1}{T}\sum\limits_i {} {h_{u,b,i}\mathbf{a}_{u,b,i}}\sum\limits_{m''}{{X}}_{u,p1}^{TF}[n',m'']\hfill \\
  {e^{ - j2\pi m''\Delta f{\tau _{u,b,i}}}}{e^{ - j2\pi {\nu _{u,b,i}}{\tau _{u,b,i}}}}{e^{j2\pi {\nu _{u,b,i}}n'T}}\hfill \\
  \int_{{\tau _{u,b,i}}}^T {} {e^{ - j2\pi \Delta ft(m' - m'' - \frac{{{\nu _{u,b,i}}}}{{\Delta f}})}}dt + \mathbf{n}_{u,b,p1}^{TF}[n',m']. \hfill \\
\end{gathered}
\end{equation}
\end{small}
\begin{IEEEproof}[Proof of (\ref{Yub1})]
  Please refer to Appendix C in \cite{hu2025novelmassiverandomaccess}.
\end{IEEEproof}
 By comparing equations (\ref{Yub1}) and (\ref{Yubp1}), it is apparent that for $0 \le n' < N'$ and $0 \le m' < M'$, define ${\mathbf{y'}}_{u,b,1}^{TF}[n',m'] = {\sqrt {\alpha \beta } {\mathbf{y}}_{u,b,1}^{TF}[\frac{{n'}}{\alpha },m']} \in {\mathbb{C}^{N_z N_y\times 1}}$ can be approximated as the received signal of ${\mathbf{X}}_{u,p1}^{TF}[n',m']$ through a channel with the same parameters, except that the Doppler parameter is $\frac{1}{\alpha }$ times the original one. Therefore, the maximum Doppler quantization parameter ${k'_{\max }}$ also becomes $\frac{1}{\alpha }$ times the original value. Based on this inference, we apply an $N' \times M'$ SFFT to each column of ${\mathbf{Y'}}_{u,b,1}^{TF}=\left [ \mathbf{y'}_{u,b,1}^{TF}\left[ {0,0} \right],...,\mathbf{y'}_{u,b,1}^{TF}\left[ {N'-1,M'-1} \right]\right]^T$ and then perform rough AUD with the maximum Doppler quantization parameter ${k'_{\max }} = [\frac{1}{\alpha }N'\Delta f{\nu _{\max }}]_{\text{R}}$. In the multi-UE scenario, the reception model for preamble1, as described in (\ref{ybp1}), can be written as:
\begin{small}
\begin{equation}
{\mathbf{Y}}_b^{p1} \approx {{\mathbf{A}}^{p1}}{\mathbf{H}}_b^{DD1} + {\mathbf{N}}_b^{DD} \in {\mathbb{C}^{N'M' \times N_zN_y}},
\end{equation}
\end{small}
where ${\mathbf{Y}}_b^{p1}$ is the SFFT result of $\sum _{u}{\mathbf{Y'}}_{u,b,1}^{TF}$, and
\begin{small}
\begin{subequations}
\begin{align}
  &\qquad \qquad \qquad {{\mathbf{A}}^{p1}} = \left[{\mathbf{A}}_1^{p1},{\mathbf{A}}_2^{p1}, \ldots ,{\mathbf{A}}_U^{p1}\right],\\
&{\mathbf{H}}_b^{DD1} = {\left[ {{{\left( {{\mathbf{H}}_{b,1}^{DD1}} \right)}^T},{{\left( {{\mathbf{H}}_{b,2}^{DD1}} \right)}^T}, \ldots ,{{\left( {{\mathbf{H}}_{b,U}^{DD1}} \right)}^T}} \right]^T}.
\end{align}
\end{subequations}
\end{small}
After completing the rough AUD, each AP transmits the detected results, representing the set of active UEs, to CPU. The CPU merges these results to form a system-wide rough active UEs set, as ${\mathcal{\bar U}_a} = \bigcup\nolimits_b {{\mathcal{\bar U}_{b,a}}} $, where ${\mathcal{\bar U}_{b,a}}$ denotes the set of active UEs detected by the $b$-th AP. Assuming that for $1 \le i \le \left| {{\mathcal{\bar U}_a}} \right|$, we have $u_i \in \mathcal{\bar U}_a$. Similarly, for multi-UE scenario, equation (\ref{ybp2}) is rewritten as:
\begin{small}
\begin{equation}
{\mathbf{Y}}_b^{p2} \approx {{\mathbf{A}}^{p2}}{\mathbf{H}}_b^{DD2} + {\mathbf{N}}_b^{DD2} \in {\mathbb{C}^{{N_p}{M_p} \times N_zN_y}},
\end{equation}
\end{small}
where ${\mathbf{Y}}_b^{p2} = \sum_u {\mathbf{Y}}_{u,b}^{p2}$, and 
\begin{small}
\begin{subequations}
\begin{align}
&\qquad \qquad \qquad {{\mathbf{A}}^{p2}} = \left[{\mathbf{A}}_{{u_1}}^{p2},{\mathbf{A}}_{{u_2}}^{p2}, \ldots ,{\mathbf{A}}_{{u_{\left| {{\mathcal{\bar U}_a}} \right|}}}^{p2}\right], \\
&{\mathbf{H}}_b^{DD2} = {\left[ {{{\left( {{\mathbf{H}}_{b,{u_1}}^{DD2}} \right)}^T},{{\left( {{\mathbf{H}}_{b,{u_2}}^{DD2}} \right)}^T}, \ldots ,{{\left( {{\mathbf{H}}_{b,{u_{\left| {{\mathcal{U}_a}} \right|}}}^{DD2}} \right)}^T}} \right]^T}.
\end{align}
\end{subequations}
\end{small}
Since $\left| {{\mathcal{\bar U}_a}} \right| \ll U$, the dimension of the sparse vector to be recovered is smaller than that of the estimated vector in a scheme that solely performs accurate AUD with the same sparsity (i.e., the same number of non-zero elements) and received signals. Therefore, a more accurate estimation can be achieved by the hybrid preamble scheme.

After obtaining active UEs and their corresponding channels, the influence of preamble1 on the received signal can be removed by using successive interference cancellation (SIC). Based on the residual received signal and estimated channel parameters, the data signal can be recovered using algorithms such as message passing. The system's signal processing flow can be seen in Fig. \ref{proc}. This part of data recovery is out of the scope of this paper. The hybrid preamble based AUD and CE scheme is summarized as in Algorithm \ref{ag2}.

\begin{algorithm}
  \caption{Hybrid Preamble Based AUD and CE Scheme}
  \label{ag2}
  \begin{algorithmic}[1]
    \REQUIRE $\left\{ {{\mathbf{Y}}_b^{p1}} \right\}$, ${{\mathbf{A}}^{p1}}$, $\left\{ {{\mathbf{Y}}_b^{p2}} \right\}$
     \ENSURE ${{ \mathcal{U}_a}} = \bigcup\nolimits_b {{{ {\mathcal{U}}}_{b,a}}} $, $\left\{ {{\mathbf{\bar H}}_b^{DD2}|1 \le b \le B} \right\}$
     \STATE {\bf \% Rough AUD}
     \FOR{$b=1$ to $B$}
     \STATE Recover ${\mathbf{H}}_{b}^{DD1}$ based on ${\mathbf{Y}}_{b}^{p1}$ and ${\mathbf{A}}_{}^{p1}$ by block sparse matrix recovery algorithm (such as GAMP-PCSBL-La proposed in Section IV);
     \STATE Obtain ${\mathcal{\bar U}_{b,a}}$ based on non-zero entries of estimated ${\mathbf{H}}_{b}^{DD1}$;
     \ENDFOR
     \STATE Form ${{\mathbf{A}}^{p2}}$ based on ${\mathcal{\bar U}_a} = \bigcup\nolimits_b {{\mathcal{\bar U}_{b,a}}} $;
     \STATE {\bf \% Accurate AUD and CE}
     \FOR{$b=1$ to $B$}
     \STATE Recover ${\mathbf{H}}_{b}^{DD2}$ based on ${\mathbf{A}}_{}^{p2}$ and ${\mathbf{Y}}_{b}^{p2}$ by block sparse matrix recovery algorithm (such as GAMP-PCSBL-La proposed in Section IV);
     \STATE Obtain accurate detected active UEs' set ${\bar {\mathcal{U}}_a}$ and channel matrix $\left\{ {{\mathbf{\bar H}}_{b,i}^{DD2}|i \in {{\bar {\mathcal{U}}}_a}} \right\}$ based on non-zero entries of estimated ${\mathbf{H}}_{b}^{DD2}$.
     \ENDFOR
  \end{algorithmic}
\end{algorithm}

\vspace{-1em}
\section{2-D Block Sparse Matrix Recovery}

\subsection{Probability Model}
As elaborated in \cite{babacan2009bayesian}, the Laplacian distribution, compared to Gaussian mixture distribution, can better capture the sparsity of signals after undergoing DCT and achieve more precise estimation. Given an AWGN channel model:
\begin{small}
\begin{equation}\label{awgnC}
{\mathbf{\tilde Y}} = {\mathbf{\tilde A\tilde X}} + {\mathbf{\tilde N}},
\end{equation}
\end{small}
where ${\mathbf{\tilde Y}} \in {\mathbb{C}^{L \times J}}$ is the observed matrix, ${\mathbf{\tilde A}} \in {\mathbb{C}^{L \times I}}$ is the measurement matrix (its values are known at the receiver), ${\mathbf{\tilde X}} \in {\mathbb{C}^{I \times J}}$ is the block sparse matrix to be estimated, and ${\mathbf{\tilde N}} \in {\mathbb{C}^{L \times J}}$ is the additive noise matrix. Since the Laplacian distribution is defined only for real-valued random variables, we need to convert the complex form model of equation (\ref{awgnC}) into the following real equivalent model:
\begin{small}
\begin{gather}\label{awgnR}
  {\mathbf{Y}} = {\mathbf{AX}} + {\mathbf{N}}, {\mathbf{Y}}  \buildrel \Delta \over =  \left[ {\begin{array}{*{20}{c}}
  {\mathcal{R}\left\{ {{\mathbf{\tilde Y}}} \right\}} \\
  {\mathcal{I}\left\{ {{\mathbf{\tilde Y}}} \right\}}
\end{array}} \right] \in {\mathbb{R}^{2L \times J}}, \notag\\
  {\mathbf{A}} \buildrel \Delta \over =  \left[ {\begin{array}{*{20}{c}}
  {\mathcal{R}\left\{ {{\mathbf{\tilde A}}} \right\}}&{ - \mathcal{I}\left\{ {{\mathbf{\tilde A}}} \right\}} \\
  {\mathcal{I}\left\{ {{\mathbf{\tilde A}}} \right\}}&{\mathcal{R}\left\{ {{\mathbf{\tilde A}}} \right\}}
\end{array}} \right] \in {\mathbb{R}^{2L \times 2I}}, \\
  {\mathbf{X}} \buildrel \Delta \over =  \left[ {\begin{array}{*{20}{c}}
  {\mathcal{R}\left\{ {{\mathbf{\tilde X}}} \right\}} \\
  {\mathcal{I}\left\{ {{\mathbf{\tilde X}}} \right\}}
\end{array}} \right] \in {\mathbb{R}^{2I \times J}}, {\mathbf{N}} \buildrel \Delta \over =  \left[ {\begin{array}{*{20}{c}}
  {\mathcal{R}\left\{ {{\mathbf{\tilde N}}} \right\}} \\
  {\mathcal{I}\left\{ {{\mathbf{\tilde N}}} \right\}}
\end{array}} \right] \in {\mathbb{R}^{2L \times J}}.\notag
\end{gather}
\end{small}
Here, $\mathcal{R}\left\{  \cdot  \right\}$ and $\mathcal{I}\left\{  \cdot  \right\}$ represent the operations of taking the real and imaginary parts of a complex matrix, respectively. In practical systems, the noise variance is often unpredictable. We assume that the communication between the transmitter and receiver occurs over an AWGN channel, i.e.,
\begin{small}
\begin{equation}
p({\mathbf{Y}}|{\mathbf{Z}}) = \prod\nolimits_{l,j} {\mathcal{N}\left( {{y_{l,j}};{z_{l,j}},\gamma } \right)},
\end{equation}
\end{small}
where ${z_{l,j}}$ is $(l,j)$-th element of matirx ${\mathbf{Z}}$, ${\mathbf{Z}} = {\mathbf{AX}}$ and $\gamma $ denotes the noise variance. Referencing the two-layer hierarchical probabilistic model of PCSBL \cite{fang2014pattern}, we introduce the hyperparameters $\left\{ {{\alpha _{i,j}}} \right\}$ and establish the probability distributions of ${\mathbf{X}}$ as:
\begin{small}
\begin{subequations}
\begin{gather}
  p({\mathbf{X}}|{\bm{\alpha }}) = {\prod\limits_{\substack{ 0 < i < I + 1 \\ j }  }{\mathcal{L}\mathcal{A}\left( {{x_{i,j}};\tau _{i,j}^{ - 1}} \right)} }  {\prod\limits_{\substack{K < i < 2I\\j}} {\mathcal{L}\mathcal{A}\left( {{x_{i,j}};\tau _{i - I,j}^{ - 1}} \right)} }, \label{pxa}\\
  {\tau _{i,j}} = {\alpha _{i,j}} + \eta  {\alpha _{i - 1,j}} + \eta  {\alpha _{i + 1,j}} + \eta  {\alpha _{i,j - 1}} + \eta  {\alpha _{i,j + 1}},\label{tauij}\\
  p({\alpha _{i,j}}) = \mathcal{G}\mathcal{A}\left( {{\alpha _{i,j}};a,b} \right). \label{paf}
\end{gather}
\end{subequations}
\end{small}

Among (\ref{pxa})-(\ref{paf}), $\mathcal{G}\mathcal{A}\left( {{\alpha _{i,j}};a,b} \right) = \Gamma {\left( a \right)^{ - 1}}{b^a}\alpha _{i,j}^a{e^{ - b{\alpha _{i,j}}}}$ denotes the Gamma distribution with shape parameter $a$ and scale parameter $b$. $\Gamma \left( a \right) = \int_0^\infty  {{t^{a - 1}}{e^{ - t}}dt} $ is the Gamma funcation. $\eta \ge 0$ represents the coupling factor and $\mathcal{L}\mathcal{A}(x;b) = \frac{1}{{2b}}\exp \left( { - \frac{{\left| x \right|}}{{2b}}} \right)$ is a Laplacian distribution. (\ref{pxa}) shows that the real and imaginary parts of ${\tilde x_{i,j}}$ share the same hyperparameter ${\tau _{i,j}}$, as defined in (\ref{tauij}).
%
\subsection{GAMP Algorithm for Sparse Matrix Recovery}
As \cite{rangan2011generalized} explained, given a prior distribution, the GAMP algorithm can achieve sparse signal recovery with reduced computational complexity from $\mathcal{O}(I^3)$ to $\mathcal{O}(IL)$. Following the sum-product and max-sum forms of the BP algorithm, the GAMP algorithm uses Gaussian and quadratic approximations to provide the minimum mean square error (MMSE) estimation and maximum a posteriori (MAP) estimation of the sparse matrix, respectively. By defining scalar estimation functions, ${g_{in}}\left(  \cdot  \right)$ and ${g_{out}}\left(  \cdot  \right)$, the GAMP algorithm iteratively performs scalar operations at the input and output nodes to decompose the vector-valued estimation problem. Assuming that in $t$-th iteration, the prior distribution of the sparse matrix is expressed as $p({\mathbf{X}}|{{\bm{\alpha }}{(t)}})$, with ${{\bm{\alpha }}{(t)}}$ is the hyperparameter obtained in the $t$-th iteration. For AWGN channel, the GAMP algorithm is shown from line 3 to line 13 in Algorithm \ref{ag1}.
\begin{algorithm} 
  \caption{GAMP-PCSBL-La} 
  \label{ag1} 
  \begin{algorithmic}[1] 
    \REQUIRE ${\mathbf{Y}}$, ${\mathbf{A}}$, $p({\mathbf{X}}|{\bm{\alpha }})$, $p({\bm{\alpha }})$, $\eta $, $ \varepsilon $ 
     \ENSURE ${\mathbf{\hat X}}\left( {t + 1} \right)$, ${\bm{\alpha }}(t + 1)$, and $\gamma (t + 1)$
     \STATE {\bf Initialize:} ${\bm{\alpha }}(1)$, ${\mathbf{\hat X}}\left(1 \right)=\mathbf{0}$, ${\mathbf{S}}\left( 0 \right)=\mathbf{0}$, $\gamma (1)$, ${u_{i,j}^x(1)}$;
     \FOR{$t=1$ to $T$}
     \STATE $\forall l, j$, $u_{l,j}^p(t) = \sum\nolimits_i {{{\left| {{a_{l,i}}} \right|}^2}} u_{i,j}^x(t)$
     \STATE $\forall l, j$, ${\hat p_{l,j}}(t) = \sum\nolimits_i {{a_{l,i}}{{\hat x}_{i,j}}(t)}  - u_{l,j}^p(t){\hat s_{l,j}}(t - 1)$
     \STATE $\forall l, j$, $u_{l,j}^z(t) = \frac{{u_{l,j}^p(t)\gamma (t)}}{{u_{l,j}^p(t) + \gamma (t)}}$
     \STATE $\forall l, j$, ${\hat z_{l,j}}(t) = \frac{{u_{l,j}^p(t){y_{l,j}} + \gamma (t){{\hat p}_{l,j}}(t)}}{{u_{l,j}^p(t) + \gamma (t)}}$
     \STATE $\forall l, j$, ${\hat s_{l,j}}(t) = {g_{out}}\left( {t,{{\hat p}_{l,j}}(t),{y_{l,j}},u_{l,j}^p(t)} \right)$
     \STATE $\forall l, j$, $u_{l,j}^s(t) =  - \frac{{\partial {g_{out}}\left( {t,{{\hat p}_{l,j}}(t),{y_{l,j}},u_{l,j}^p(t)} \right)}}{{\partial {{\hat p}_{l,j}}(t)}}$
     \STATE $\forall i, j$, $u_{i,j}^r(t) = {\left[ {\sum\nolimits_l {{{\left| {{a_{l,i}}} \right|}^2}u_{l,j}^s(t)} } \right]^{ - 1}}$
     \STATE $\forall i, j$, ${\hat r_{i,j}}(t) = {\hat x_{i,j}}(t) + u_{i,j}^r(t)\sum\nolimits_l {{a_{l,i}}{{\hat s}_{l,j}}(t)}$
     \STATE $\forall i, j$, ${\tau _{i,j}}(t) = {\alpha _{i,j}}(t) + \eta  {\alpha _{i - 1,j}}(t) + \eta  {\alpha _{i + 1,j}}(t) + \eta  {\alpha _{i,j - 1}}(t) + \eta  {\alpha _{i,j + 1}}(t)$
     \STATE $\forall i, j$, ${\hat x_{i,j}}(t + 1) = {g_{in}}\left( {t,{{\hat r}_{i,j}}(t),{\tau _{i,j}}(t),u_{i,j}^r(t)} \right)$
     \STATE $\forall i, j$, $u_{i,j}^x(t + 1) = u_{i,j}^r(t)\frac{{\partial {g_{in}}\left( {t,{{\hat r}_{i,j}}(t),{\tau _{i,j}}(t),u_{i,j}^r(t)} \right)}}{{\partial {{\hat r}_{i,j}}(t)}}$
     \STATE $\forall i, j$, ${\alpha _{i,j}}(t + 1) = \frac{a}{{b + {\omega _{i,j}}(t + 1) + {\omega _{N + i,j}}(t + 1)}}$
     \STATE Update $\gamma (t + 1) = \frac{{\sum\nolimits_{l,j} {} {{\left\| {{y_{l,j}} - {{\hat z}_{l,j}}(t)} \right\|}^2} + u_{l,j}^z(t)}}{{2MN}}$
     \STATE If $\frac{{\left\| {{\mathbf{\hat X}}\left( {t + 1} \right) - {\mathbf{\hat X}}\left( t \right)} \right\|_F^2}}{{\left\| {{\mathbf{\hat X}}\left( {t + 1} \right)} \right\|_F^2}} < \varepsilon $, break
     \ENDFOR
  \end{algorithmic}
\end{algorithm}

For sum-product GAMP, ${g_{out}}\left( {t,{{\hat p}_{l,j}}(t),{y_{l,j}},u_{l,j}^p(t)} \right)$ and $u_{l,j}^s(t)$ are defined as
\begin{small}
\begin{subequations}
\begin{gather}
{g_{out}}\left( {t,{{\hat p}_{l,j}}(t),{y_{l,j}},u_{l,j}^p(t)} \right) = \frac{{{y_{l,j}} - {{\hat p}_{l,j}}(t)}}{{u_{l,j}^p(t) + \gamma (t)}},\\
u_{l,j}^s(t) = \frac{1}{{u_{l,j}^p(t) + \gamma (t)}}.
\end{gather}
\end{subequations}
\end{small}
Based on MMSE estimation, in input node, we have
\begin{small}
\begin{subequations}
\begin{align}
{g_{in}}\left( {t,{{\hat r}_{i,j}}(t),{\tau _{i,j}}(t),u_{i,j}^r(t)} \right) &= {\mathbb{E}}_{p(x|r,\tau,u^r)}\{ {x_{i,j}} \},\\
\frac{{\partial {g_{in}}\left( {t,{{\hat r}_{i,j}}(t),{\tau _{i,j}}(t),u_{i,j}^r(t)} \right)}}{{\partial {{\hat r}_{i,j}}(t)}}{\tau _{i,j}}(t) &= {\mathbb{V}}_{p(x|r,\tau,u^r)}\{ {x_{i,j}} \},
\end{align}
\end{subequations}
\end{small}
where $p(x_{i,j}|r,\tau,u^r)$ represent the approximate posterior distribution of $(i,j)$-th element of the matrix to be estimated. In the sum-product derivation, the messages from the factor node $p\left( {y|x} \right)$ to the variable node ${x_{i,j}}$ are approximated as:
\begin{small}
\begin{equation}
{\vec m_{{x_{i,j}}}}(t) \approx \mathcal{N}\left( {{x_{i,j}};{{\hat r}_{i,j}}(t),u_{i,j}^r(t)} \right).
\end{equation}
\end{small}
As previously mentioned, the prior of ${x_{i,j}}$ is defined as:
\begin{small}
\begin{align*}
p({x_{i,j}}|{\tau _{\tilde i,j}}(t)) = \mathcal{L}\mathcal{A}\left( {{x_{i,j}};{{\left( {\tau _{\tilde i,j}^{}(t)} \right)}^{ - 1}}} \right),
\end{align*}
\end{small}
with
\begin{small}
\begin{align*}
\tilde i = \left\{ {\begin{array}{*{20}{c}}
  i&{1 \le i \le I} \\
  {i - I}&{I + 1 \le i \le 2I}
\end{array}} \right..
\end{align*}
\end{small}
Therefore, the approximate posterior distribution of ${x_{i,j}}$ can be expressed as:
\begin{small}
\begin{align}\label{78}
  &p\left( {{x_{i,j}}|{{\hat r}_{i,j}}(t),{\tau _{i,j}}(t),u_{i,j}^r(t)} \right) \propto {{\vec m}_{{x_{i,j}}}}(t)p({x_{i,j}}|{\tau _{\tilde i,j}}(t)) \notag \hfill \\
  & = \mathcal{N}\left( {{x_{i,j}};{{\hat r}_{i,j}}(t),u_{i,j}^r(t)} \right)\mathcal{L}\mathcal{A}\left( {{x_{i,j}};{{\left( {\tau _{\tilde i,j}^{}(t)} \right)}^{ - 1}}} \right) \hfill \\
  & = \frac{1}{{{\psi _{i,j}}(t)}}\exp \left\{ { - {\xi _{i,j}}\left( t \right) - \frac{1}{{2u_{i,j}^r(t)}}{{\left( {{x_{i,j}} - {\varphi _{i,j}}\left( t \right)} \right)}^2}} \right\},\notag
\end{align}
\end{small}
where
\begin{small}
\begin{subequations}
\begin{align}\label{79}
  &{\psi _{i,j}}(t) = \sqrt {2\pi u_{i,j}^r(t)} \left[ \exp \left\{ { - \xi _{i,j}^ - \left( t \right)} \right\}Q\left( {{\varphi _{i,j}^ - (t)}}/{{\sqrt {u_{i,j}^r(t)} }} \right)\right. \notag \\
  & \left. + \exp \left\{ { - \xi _{i,j}^ + \left( t \right)} \right\}Q\left( {{{\varphi _{i,j}^ + (t)}}/{{\sqrt {u_{i,j}^r(t)} }}} \right) \right],\\
&{\xi _{i,j}}\left( t \right) = {\tau _{\tilde i,j}}(t){\hat r_{i,j}}(t){\text{sign}}\left( {{x_{i,j}}} \right) - \frac{1}{2}u_{i,j}^r(t){\left( {{\tau _{\tilde i,j}}(t)} \right)^2},\\
&{\varphi _{i,j}}\left( t \right) = {\hat r_{i,j}}(t) - u_{i,j}^r(t){\tau _{\tilde i,j}}(t){\text{sign}}\left( {{x_{i,j}}} \right),\\
&\xi _{i,j}^ - \left( t \right) =  - {\tau _{\tilde i,j}}(t){\hat r_{i,j}}(t) - \frac{1}{2}u_{i,j}^r(t){\left( {{\tau _{\tilde i,j}}(t)} \right)^2},\label{82}\\
&\xi _{i,j}^ + \left( t \right) = {\tau _{\tilde i,j}}(t){\hat r_{i,j}}(t) - \frac{1}{2}u_{i,j}^r(t){\left( {{\tau _{\tilde i,j}}(t)} \right)^2},\label{83}\\
&\varphi _{i,j}^ - \left( t \right) = {\hat r_{i,j}}(t) + u_{i,j}^r(t){\tau _{\tilde i,j}}(t),\label{84}\\
&\varphi _{i,j}^ + \left( t \right) = {\hat r_{i,j}}(t) - u_{i,j}^r(t){\tau _{\tilde i,j}}(t),\label{85}\\
&{\text{sign}}({x_{i,j}}) = \left\{ {\begin{array}{*{20}{c}}
  1,&{{x_{i,j}} > 0,} \\
  0,&{{x_{i,j}} = 0,} \\
  { - 1},&{{x_{i,j}} < 0.}
\end{array}} \right.
\end{align}
\end{subequations}
\end{small}
According to (\ref{78}), the posterior mean and variance of ${x_{i,j}}$ can be calculated as in (\ref{hatxij}) and (\ref{uijx}), where $Q\left(  \cdot  \right)$ is the standard $Q$-function, representing the tail probability of the normal distribution, defined as:
\begin{figure*}[htbp]
  \vspace{-1.9em}
  \centering
  \begin{small}
    \begin{equation}\label{hatxij}
    {\hat x_{i,j}}(t + 1) = \frac{{\sqrt {2\pi u_{i,j}^r(t)} }}{{{\psi _{i,j}}(t)}}\left[ {{e^{ - \xi _{i,j}^ - (t)}}\varphi _{i,j}^ - (t)Q\left( {{{\varphi _{i,j}^ - (t)}}/{{\sqrt {u_{i,j}^r(t)} }}} \right) + {e^{ - \xi _{i,j}^ + (t)}}\varphi _{i,j}^ + (t)Q\left( {{{\varphi _{i,j}^ + (t)}}/{{\sqrt {u_{i,j}^r(t)} }}} \right)} \right].
    \end{equation}
    \end{small}
    \vspace{-2.5em}
\end{figure*}
\begin{figure*}[htbp]
  \centering
  \begin{small}
    \begin{equation}\label{uijx}
    \begin{gathered}
    u_{i,j}^x(t + 1) = \frac{{\sqrt {2\pi u_{i,j}^r(t)} }}{{{\psi _{i,j}}(t)}}\left( \left( {{{\left( {\varphi _{i,j}^ + (t)} \right)}^2} + u_{i,j}^r(t)} \right){e^{ - \xi _{i,j}^ + (t)}}Q\left( { - {{\varphi _{i,j}^ + (t)}}/{{\sqrt {u_{i,j}^r(t)} }}} \right) + \left( {{{\left( {\varphi _{i,j}^ - (t)} \right)}^2} + u_{i,j}^r(t)} \right){e^{ - \xi _{i,j}^ - (t)}} \right. \hfill \\
    Q\left( {{{\varphi _{i,j}^ - (t)}}/{{\sqrt {u_{i,j}^r(t)} }}} \right)
    \left. - \frac{{2{\tau _{i,j}}(t)u_{i,j}^r{{(t)}^2}}}{{\sqrt {2\pi u_{i,j}^r(t)} }}{e^{ - {{{{\hat r}_{i,j}}{{(t)}^2}}}/{{2u_{i,j}^r(t)}}}} \right)
    - {\left( {{{\hat x}_{i,j}}(t)} \right)^2}. \hfill \\
    \end{gathered}
    \end{equation}
    \end{small}
    \hrulefill
    \vspace{-1.8em}
\end{figure*}
\begin{small}
\begin{equation}
Q\left( x \right) = \frac{1}{{\sqrt {2\pi } }}\int_x^\infty  {{e^{ - \frac{{{u^2}}}{2}}}du}.
\end{equation}
\end{small}
\begin{IEEEproof}[\hspace{-2.5em} Derivation of (\ref{hatxij}) and (\ref{uijx})]
Please refer to Appendix D in \cite{hu2025novelmassiverandomaccess}.
\end{IEEEproof}
This completes the GAMP portion of Algorithm \ref{ag1}.
\vspace{-1.1em}

\subsection{Learning Hyperparameters via EM Algorithm}
After obtaining the posterior distribution of ${\mathbf{X}}$, our objective shifts to finding appropriate hyperparameters ${\bm{\alpha }}$ and $\gamma $ that maximize the posterior probability of them. A direct strategy is to use the EM algorithm, where ${\mathbf{X}}$ is treated as a hidden variable. In the E-step, the log-posterior mean is computed, and in the M-step, the log-posterior is maximized. The iterative process of these two steps is summarized as follows.

\textbf{\textit{E Step:}} Given the posterior distribution of ${\mathbf{X}}$ and the observed matrix ${\mathbf{Y}}$, we compute the mean of the log-posterior of the hyperparameters ${\bm{\alpha }}$ with respect to the hidden variable ${\mathbf{X}}$ in $t$-th iteration. Let ${\bm{\theta }} = \left\{ {{\bm{\alpha }},\gamma } \right\}$ and we define $R$ function as:
\begin{small}
\begin{equation}
\begin{gathered}
  R\left( {{\bm{\theta }}|{\bm{\theta }}(t)} \right) = {\mathbb{E}_{p({\mathbf{X}}|{\mathbf{Y}},{{\bm{\theta }}{\left( t \right)}})}}\left\{ {\log p\left( {{\bm{\theta }}|{\mathbf{X}},{\mathbf{Y}},{\bm{\theta }}(t)} \right)} \right\} \hfill \\
   = R\left( {{\bm{\alpha }}|{\bm{\theta }}(t)} \right) + R\left( {\gamma |{\bm{\theta }}(t)} \right) + c, \hfill \\
\end{gathered}
\end{equation}
\end{small}
where $c$ represents a constant that is independent of ${\bm{\theta }}$. Next, we calculate $R\left( {{\bm{\alpha }}|{\bm{\theta }}(t)} \right)$ and $R\left( {\gamma |{\bm{\theta }}(t)} \right)$ as follow.
\begin{small}
\begin{align}
  &R\left( {{\bm{\alpha }}|{\bm{\theta }}(t)} \right) = {\mathbb{E}_{p({\mathbf{X}}|{\mathbf{Y}},{\bm{\theta }}(t))}}\left\{ {\log p\left( {{\mathbf{X}}|{\bm{\alpha }}} \right) + \log p\left( {\bm{\alpha }} \right)} \right\} \notag \hfill \\
  & = \sum\nolimits_{i,j} {} 2\ln ({\alpha _{i,j}} + \eta  {\alpha _{i - 1,j}} + \eta  {\alpha _{i + 1,j}} + \eta  {\alpha _{i,j - 1}} + \eta  {\alpha _{i,j + 1}}) \notag \hfill \\
  & - ({\alpha _{i,j}} + \eta  {\alpha _{i - 1,j}} + \eta  {\alpha _{i + 1,j}} + \eta  {\alpha _{i,j - 1}} + \eta  {\alpha _{i,j + 1}}) \times \notag \hfill \\
  &\left\langle {\left| {{x_{i,j}}(t)} \right| + \left| {{x_{i + I,j}}(t)} \right|} \right\rangle  + a\ln {\alpha _{i,j}} - b{\alpha _{i,j}}, 
\end{align}
\end{small}
\begin{small}
\begin{equation}\label{Rgamma}
\begin{gathered}
  R\left( {\gamma |{\mathbf{\theta }}(t)} \right) = {\mathbb{E}_{p({\mathbf{X}}|{\mathbf{Y}},{\mathbf{\theta }}(t))}}\left\{ {\log p\left( {{\mathbf{Y}}|{\mathbf{Z}},\gamma } \right)} \right\} \hfill \\
   =  - IJ\ln \gamma  - \frac{{\left\| {{\mathbf{Y}} - {\mathbf{\hat Z}}\left( t \right)} \right\|_F^2 + \sum\nolimits_{i,j} {u_{i,j}^z(t)} }}{{2\gamma }},\hfill \\
\end{gathered}
\end{equation}
\end{small}
where $\left\langle {\left| {{x_{i,j}}(t)} \right|} \right\rangle $ represents the mean of the absolute value of ${x_{i,j}}(t)$, that is
\begin{small}
\begin{equation}
\begin{gathered}
  \left\langle {\left| {{x_{i,j}}(t)} \right|} \right\rangle  = \int {\left| {{x_{i,j}}(t)} \right|p({x_{i,j}}(t)|{\mathbf{Y}},{\mathbf{\theta }}(t))d} {x_{i,j}}(t) \hfill \\
   = \frac{1}{{{\psi _{i,j}}(t)}}\left[ {e^{ - \xi _{i,j}^ + (t)}}\sqrt {2\pi u_{i,j}^r(t)} \varphi _{i,j}^ + (t)Q\left( {\frac{{\varphi _{i,j}^ + (t)}}{{\sqrt {u_{i,j}^r(t)} }}} \right) - \right. \hfill \\
   \left. {e^{ - \xi _{i,j}^ - (t)}}\sqrt {2\pi u_{i,j}^r(t)} \varphi _{i,j}^ - (t)Q\left( {\frac{{\varphi _{i,j}^ - (t)}}{{\sqrt {u_{i,j}^r(t)} }}} \right) + 2u_{i,j}^r(t){e^{ - \frac{{{{\left( {{{\hat r}_{i,j}}(t)} \right)}^2}}}{{2u_{i,j}^r(t)}}}} \right]. \hfill \\
\end{gathered}
\end{equation}
\end{small}
\textbf{\textit{M Step: }}We update the hyperparameters ${\bm{\alpha }}$ and $\gamma$ by maximizing $R$ function:
\begin{small}
\begin{subequations}
\begin{align}
{\bm{\alpha }}(t + 1) &= \mathop {\arg \max }\limits_{\bm{\alpha }} R\left( {{\bm{\alpha }}|{\bm{\theta }}(t)} \right),\label{alphat}\\
\gamma (t + 1) &= \mathop {\arg \max }\limits_\gamma  R\left( {\gamma |{\bm{\theta }}(t)} \right).\label{gammat}
\end{align}
\end{subequations}
\end{small}
First, we consider $\bm{\alpha}$. Unlike conventional SBL, in PCSBL, the hyperparameters are interdependent, meaning that the element-wise estimation of parameters cannot be performed independently. Directly solving the result of (\ref{alphat}) is challenging. To address this, we refer to the derivation process in \cite{bellili2019generalized} and consider an alternative suboptimal solution that achieves good estimation accuracy while simplifying the computation process. Assuming ${{\bm{\alpha }}^*}$ is the optimal solution to (\ref{alphat}), the first-order derivative of R funcation with respect to ${\bm{\alpha }}$ equals zero at ${{\bm{\alpha }}^*}$. That is, for any $i$, $j$, the following condition holds:
\begin{small}
\begin{equation}\label{condi}
\begin{gathered}
  \left. \frac{{\partial R( {{\mathbf{\alpha }}|{{\mathbf{\theta }}^{(t)}}})}}{{\partial {\alpha _{i,j}}}} \right|_{{\mathbf{\alpha }} = {{\mathbf{\alpha }}^*}} = \frac{a}{{\alpha _{i,j}^*}} + 2\left( {\upsilon _{i,j}} + \eta  {\upsilon _{i - 1,j}} + \eta  {\upsilon _{i + 1,j}} + \right. \hfill \\
  \left. \eta  {\upsilon _{i,j - 1}} +\eta  {\upsilon _{i,j + 1}} \right) - b - {\omega _{i,j}}(t) - {\omega _{N + i,j}}(t) = 0, \hfill \\
\end{gathered}
\end{equation}
\end{small}
where
\begin{small}
\begin{equation}
\begin{gathered}
  {\omega _{i,j}}(t) \buildrel \Delta \over = \left\langle {\left| {{x_{i,j}}(t)} \right|} \right\rangle  + \eta  \left\langle {\left| {{x_{i - 1,j}}(t)} \right|} \right\rangle  \hfill \\
   + \eta  \left\langle {\left| {{x_{i + 1,j}}(t)} \right|} \right\rangle  + \eta  \left\langle {\left| {{x_{i,j - 1}}(t)} \right|} \right\rangle  + \eta  \left\langle {\left| {{x_{i,j + 1}}(t)} \right|} \right\rangle, \hfill \\
\end{gathered}
\end{equation}
\end{small}
\begin{small}
\begin{equation}\label{vij}
{\upsilon _{i,j}} \buildrel \Delta \over = \frac{1}{{\alpha _{i,j}^* + \eta  \alpha _{i - 1,j}^* + \eta  \alpha _{i + 1,j}^* + \eta  \alpha _{i,j - 1}^* + \eta  \alpha _{i,j + 1}^*}}.
\end{equation}
\end{small}
In our model, the parameters $\eta   \ge 0$ and ${\alpha _{i,j}} \ge 0$ hold true for any $i, j$. Building on this, based on equation (\ref{vij}), ${\upsilon _{i,j}}$ satisfies the following inequality constraints:
\begin{small}
\begin{gather}\label{inequa}
  0 \le {\upsilon _{i,j}} \le \frac{1}{{\alpha _{i,j}^*}}, \notag\\
  0 \le {\upsilon _{i,j}} \le \frac{1}{{\eta  \alpha _{i - 1,j}^*}}, \notag\\
  0 \le {\upsilon _{i,j}} \le \frac{1}{{\eta  \alpha _{i + 1,j}^*}}, \\
  0 \le {\upsilon _{i,j}} \le \frac{1}{{\eta  \alpha _{i,j - 1}^*}}, \notag\\
  0 \le {\upsilon _{i,j}} \le \frac{1}{{\eta  \alpha _{i,j + 1}^*}}. \notag
\end{gather}
\end{small}
Substituting the above results into equation (\ref{condi}), we obtain:
\begin{small}
\begin{equation}
\frac{a}{{\alpha _{i,j}^*}} \le b + {\omega _{i,j}}(t) + {\omega _{N + i,j}}(t) \le \frac{{a + 10}}{{\alpha _{i,j}^*}}.
\end{equation}
\end{small}
Then $\alpha _{i,j}^* \in \left[ {\frac{a}{{b + {\omega _{i,j}}(t) + {\omega _{N + i,j}}(t)}},\frac{{a + 10}}{{b + {\omega _{i,j}}(t) + {\omega _{N + i,j}}(t)}}} \right]$ is held. Therefore, a simple suboptimal solution for equation (\ref{alphat}) can be given by:
\begin{small}
\begin{equation}\label{upalpha}
{\alpha _{i,j}}(t + 1) = \frac{a}{{b + {\omega _{i,j}}(t) + {\omega _{N + i,j}}(t)}},
\end{equation}
\end{small}

Next, we focus on noise variance $\gamma $. Suppose ${\gamma ^*}$ is the optimal solution of (\ref{gammat}), it satisfies:
\begin{small}
\begin{equation}
\left. \frac{{\partial R\left( {\gamma |{\mathbf{\theta }}(t)} \right)}}{{\partial \gamma }} \right| _{\gamma  = {\gamma ^*}} =  - \frac{{MN}}{{{\gamma ^*}}} + \frac{{\left\| {{\mathbf{Y}} - {\mathbf{\hat Z}}\left( t \right)} \right\|_F^2 + \sum\nolimits_{i,j} {u_{i,j}^z(t)} }}{{2{{\left( {{\gamma ^*}} \right)}^2}}} = 0.
\end{equation}
\end{small}
It is easy to obtain the expression of $\gamma (t + 1)$ as:
\begin{small}
\begin{equation}\label{upgamma}
\gamma (t + 1) = {\gamma ^*} = \frac{{\left\| {{\mathbf{Y}} - {\mathbf{\hat Z}}\left( t \right)} \right\|_F^2 + \sum\nolimits_{i,j} {u_{i,j}^z(t)} }}{{2MN}}.
\end{equation}
\end{small}
Thus completes the update process for ${\bm{\theta }}$. Equations (\ref{upalpha}) and (\ref{upgamma}) serve as the output of the EM algorithm, reflected in lines 14 and 15 of Algorithm \ref{ag1}. With this, we have completed the entire derivation process of the GAMP-PCSBL-La algorithm. In Section VI, we validate that the proposed GAMP-PCSBL-La algorithm can accurately estimate block sparse matrix with DCT sparse properties. This algorithm is employed for AUD and CE.

\vspace{-1em}
\section{Computational Complexity Analysis}
The scheme proposed in this paper consists of two main stages: rough AUD and joint accurate AUD and CE. For rough AUD, an additional SFFT for the superimposed preamble is introduced, along with the GAMP-PCSBL-La algorithm for 2-D block sparse matrix recovery. The computational complexity for rough AUD is given by ${\chi _s} = {\mathcal{O}}\left( {N'\log N'} \right) + {\mathcal{O}}\left( {M'\log M'} \right) + {\mathcal{O}}\left( {{N_z}{N_y}N'M'U\left| {{{\mathbf{p}}_{r1}}} \right|} \right)$, where the first two terms correspond to the SFFT, and the last term corresponds to the GAMP-PCSBL-La algorithm. Similarly, for the joint accurate AUD and CE, the GAMP-PCSBL-La algorithm for 2-D block sparse matrix recovery, applied to the embedded preamble, is used, with a computational complexity of ${\chi _e} = {\mathcal{O}}\left( {{N_z}{N_y}\left| {{{\mathcal{K}}_{\mathcal{A}}}} \right|\left| {{{\mathbf{p}}_{r2}}} \right|{N_p}{M_p}} \right)$. Additionally, the receiver needs to perform SIC for the superimposed preamble, with a computational complexity of $M'N'L$. In summary, the overall computational complexity of the proposed scheme is ${\chi _h} = {\chi _s} + {\chi _e} + M'N'L$. In existing schemes, the superimposed scheme has lower complexity but poorer access performance. Furthermore, as $U \gg \left| {{{\mathcal{K}}_{\mathcal{A}}}} \right|$, this results in the embedded scheme potentially having higher complexity than the hybird scheme, while occupying more DD resources.

\vspace{-0.5em}
\section{Simulations}
\begin{figure*}[htbp]
  \vspace{-1.9em}
	\centering
	\subfigure[]{\includegraphics[width=2.3in]{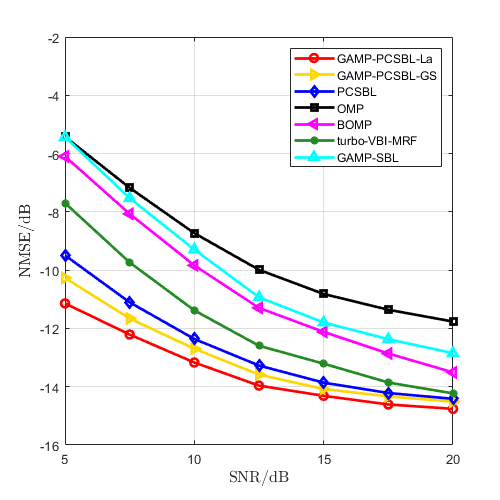}}
  \subfigure[]{\includegraphics[width=2.3in]{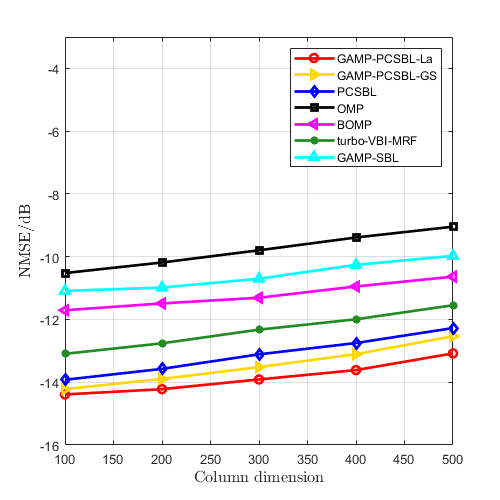}}
	\subfigure[]{\includegraphics[width=2.3in]{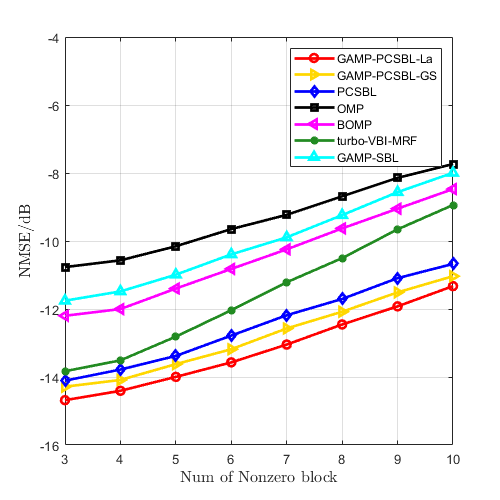}}
	\caption{Performance comparisons for block sparse matrix recovery algorithms.}
	\label{A}
  \vspace{-1.5em}
\end{figure*}
To validate the effectiveness and superiority of the proposed scheme, numerical simulations are conducted. The specific simulation parameters are detailed in Table \ref{tab1}. We consider the 3GPP vehicular models, namely extended vehicular A (EVA) with number of paths is 9 and ${\tau _{\max }} = 2.5\mu s$ \cite{series2009guidelines}. The delay and Doppler parameters are randomly generated within the range of 0 to their respective maximum values. ${\vartheta _{u,b,i}}$ and ${\varsigma _{u,b,i}}$ are uniformly distributed within the ranges $\left[ {0,\pi } \right]$ and $\left[ { - \frac{\pi }{2},\frac{\pi }{2}} \right]$, respectively.

\begin{table}[!htbp]
    \vspace{-1.5em}
    \centering
    \caption{Simulation Parameters}
    \begin{tabular}{|c|c|c|}
        \hline
        Parameters & Definition & Value\\
        \hline
        $N$ & Doppler dimension for a block & 128\\
        \hline
        $M$ & Delay dimension for a block & 512\\
        \hline
        $P$ & Number of a UE's paths & 9\\
        \hline
        $v_{\max}$ & Maximum velocity (km/h) & 300\\
        \hline
        $\tau_{\max}$& Maximum path delay ($\mu$s) & 2.5\\
        \hline
        $f_c$ & Carrier frequency (GHz) & 4\\
        \hline
        $\Delta f$ & Subcarrier interval (kHz) & 15\\
        \hline
        $N'$ & Doppler dimension for preamble1 & 64\\
        \hline
        $M'$ & Delay dimension for preamble1 & 4\\
        \hline
        $K_p$ & Doppler dimension for preamble2 & 20\\
        \hline
        $L_p$ & Delay dimension for preamble2 & 20\\
        \hline
        $\lambda$ & Large scale fading coefficient (dB) & $-128-37.6\log d$\\
        \hline
        $n_0$ & Background noise (dBm/Hz) & -174\\
        \hline
        $P_t$ & Transmission power (dBm) & 10\\
        \hline
    \end{tabular}
    
    \label{tab1}
    \vspace{-1em}
\end{table}

To evaluate the performance of the massive random access scheme, we use the detection error rate (DER) and the normalized mean squared error (NMSE) as performance metrics for AUD and CE, respectively. They are defined as follows:

\begin{small}
  \begin{subequations}
     \begin{align}
      & \qquad {\text{DER}} = \frac{{\left| {{{\mathcal{K_A}}}\backslash {{{\mathcal{U}}}_a}} \right| + \left| {{{{\mathcal{U}}}_a}\backslash {{\mathcal{K_A}}}} \right|}}{U}, \\
      &{\text{NMSE}} = 10{\log _{10}}\frac{{\left\| {{\mathbf{\bar H}}_{}^{DDA2} - {\mathbf{H}}_{}^{DDA2}} \right\|_F^2}}{{\left\| {{\mathbf{H}}_{}^{DDA2}} \right\|_F^2}},
     \end{align}
  \end{subequations}
\end{small}
where ${\mathcal{A}}\backslash {\mathcal{B}}$ represents a set whose elements are in ${\mathcal{A}}$ but not in ${\mathcal{B}}$. $\left| {\mathcal{A}} \right|$ denotes the cardinality of set $\mathcal{A}$. A smaller DER or NMSE indicates more accurate detection and estimation results, corresponding to better AUD and CE performance.

Initially, the performance of the proposed GAMP-PCSBL-La algorithm is compared with other existing algorithms for 2-D block sparse matrix recovery. we set the dimensions of the 2-D block sparse matrix to $256 \times 64$, the observation matrix to $64 \times 64$, and the sensing matrix to $64 \times 256$. A block sparse matrix was generated by randomly creating non-zero values and applying a DCT. The elements of the sensing matrix and noise matrix followed a Gaussian distribution. Compared algorithms including GAMP-PCSBL with Gaussian prior (GAMP-PCSBL-Gs) \cite{fang2016two}, PCSBL \cite{fang2014pattern}, orthogonal matching pursuit (OMP) \cite{tropp2007signal}, block OMP (BOMP) \cite{lu2022compressive}, turbo variational Bayesian inference with Markov random field (turbo-VBI-MRF) \cite{zhang2020variance}, and GAMP-SBL \cite{zhang2023generalized}. 
\begin{figure}[htb]
  \vspace{-1em}
  \begin{center}
  {\includegraphics[scale=0.5]{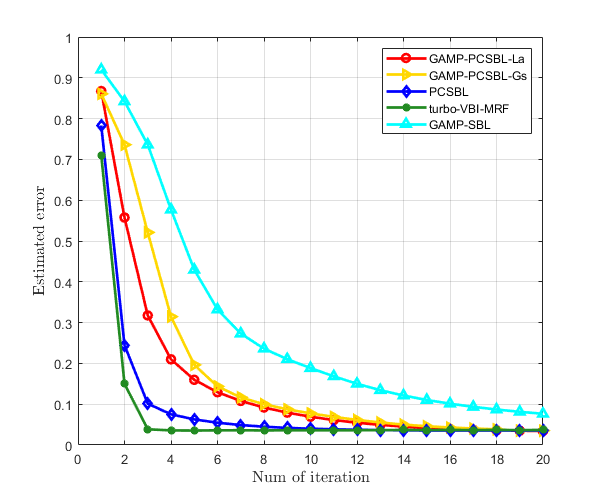}}\\
  \end{center}
  \caption{Convergence trends of iterative algorithms.}
  \label{Conv}
  \vspace{-0.5em}
\end{figure}

In Fig. \ref{A}(a), with the number of non-zero blocks fixed at 5, we compared the performance of various algorithms in recovering block sparse matrix under different signal-to-noise ratios (SNRs). The simulation curves show that as the SNR increases, the NMSE performance of all algorithms improves. In Fig. \ref{A}(b), with the SNR fixed at 12.5 dB, we analyzed the impact of varying the column dimensions of the block sparse matrix on the performance of each algorithm. It is evident that as the dimensions of the sparse matrix increase, the estimation accuracy of all algorithms gradually declines. Additionally, in Fig. \ref{A}(c), with the SNR fixed at 12.5 dB and the sparse matrix dimensions set to $256 \times 64$, we compared the performance trends of each algorithm under different numbers of non-zero blocks. This figure implies that as the number of non-zero blocks increases, the estimated accuracy decreases across all algorithms. The simulation results in Fig. \ref{A}(b) and \ref{A}(c) are consistent with the relevant conclusions of compressed sensing theory. These simulation curves also demonstrate that algorithms utilizing PCSBL outperform other algorithms in block sparse matrix recovery. Moreover, the proposed GAMP-PCSBL-La algorithm outperforms the other algorithms, showcasing its unique performance advantages in recovering block sparse matrices formed through DCT.
\begin{figure}[ht]
  \vspace{-1em}
  \begin{center}
  {\includegraphics[scale=0.45]{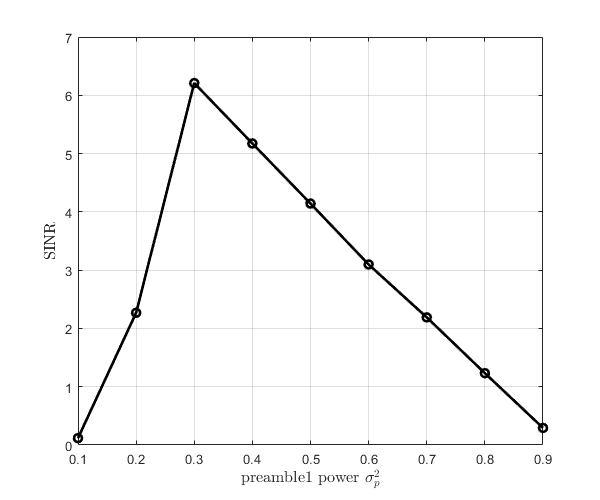}}\\
  \end{center}
  \caption{SINR versus $\sigma _p^2$ for the proposed hybrid scheme with $|\mathcal{K_A}|=30$ and $N_y=N_z=8$.}
  \label{Pow}
  \vspace{-0.5em}
\end{figure}
\begin{figure*}[hb]
  \vspace{-1.5em}
	\centering
	\subfigure[]{\includegraphics[width=3.3in]{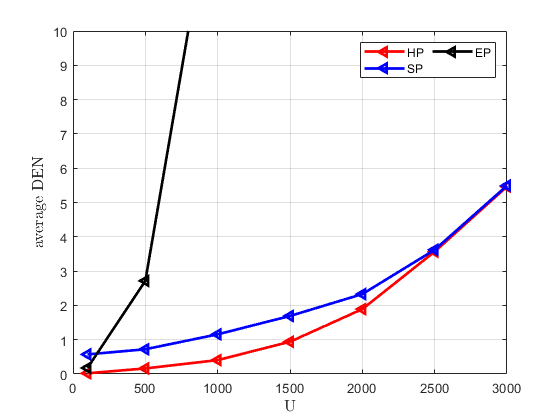}}
  \hspace{4mm}
  \subfigure[]{\includegraphics[width=3.3in]{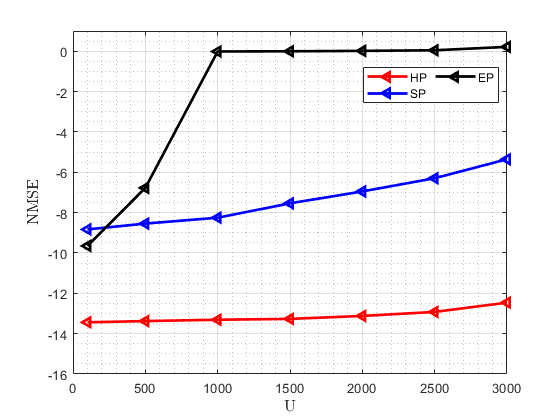}}
	\caption{Trend of different schemes' performance with varying $U$ under fixed $|\mathcal{K_A}| =30$ and $N_y = N_z = 8$: (a) average DEN for AUD; (b) NMSE for CE.}
	\label{U}
\end{figure*}

We compare the convergence trends of several iterative algorithms in Fig. \ref{Conv}, with the simulation settings being consistent with those in Fig. \ref{A}. The `estimated error' in the figure is defined as the non-logarithmic form of NMSE, i.e., $10^{\text{NMSE}/10}$. The figure shows that the PCSBL and turbo-VBI-MRF algorithms, which are based on direct matrix inversion, converge faster than GAMP-based algorithms, reaching convergence in approximately 5 iterations. In contrast, the GAMP-based algorithms converge after about 20 iterations. This simulation result demonstrates that the proposed GAMP-PCSBL-La algorithm exhibits good convergence and reliability.

In the hybrid preamble scheme proposed in this paper, the power allocation between the two superimposed signals, ${\mathbf{X}}_{u,1}^{DD}$ and ${\mathbf{X}}_{u,2}^{DD}$, significantly impacts the performance of the receiver. To assess this impact, we define $\sigma _p^2 = \frac{{{P_{{X_1}}}}}{{{P_{{X_1}}} + {P_{{X_2}}}}}$ as the power allocation ratio of ${\mathbf{X}}_{u,1}^{DD}$, and let $\sigma _p^2$ vary from 0.1 to 0.9 in steps of 0.1. After SIC, the received signal is expressed as ${{\mathbf{\hat Y}}_b} = {{\mathbf{Y}}_b} - {{\mathbf{S}}_{{\mathbf{X}}_1^{DD},{\mathbf{\hat H}}_b^{DD2}}} = {{\mathbf{S}}_{{\mathbf{X}}_2^{DD},{\mathbf{\hat H}}_b^{DD2}}} + {\mathbf{I}} + {\mathbf{N}}$, where ${{\mathbf{S}}_{{\mathbf{X}},{\mathbf{ H}}}}$ represents the received signal of ${\mathbf{X}}$ after passing through the channel ${\mathbf{H}}$. ${\mathbf{I}}$ and ${\mathbf{N}}$ denote interference and additive Gaussian noise, respectively.  The receiver performance is evaluated based on the signal-to-interference-plus-noise ratio (SINR) under different values of $\sigma _p^2$:
\begin{equation}
{\text{SINR}} = \frac{{\sum\nolimits_b {\left\| {{{\mathbf{S}}_{{\mathbf{X}}_2^{DD},{\mathbf{\hat H}}_b^{DD2}}}} \right\|_F^2} }}{{\sum\nolimits_b {\left\| {{{{\mathbf{\hat Y}}}_b} - {{\mathbf{S}}_{{\mathbf{X}}_2^{DD},{\mathbf{\hat H}}_b^{DD2}}}} \right\|_F^2} }}.
\end{equation}
The simulation results, as shown by the curves in Fig.\ref{Pow}, indicate that as $\sigma _p^2$ increases, the receiver performance first improves and then deteriorates, with optimal performance occurring around $\sigma _p^2=0.3$. At this point,  the magnitude of non-zero entries of ${\mathbf{X}}_{u,1}^{DD}$ to be ten times that of ${\mathbf{X}}_{u,2}^{DD}$. The trend of this curve aligns with the conclusion in \cite{mishra2021otfs}, which indicates that there is an optimal power allocation ratio for the superimposed pilot signals. In the subsequent simulations, we will fix the $\sigma _p^2=0.3$.

In order to demonstrate the advantages of the proposed scheme in massive random access scenarios, we compare its performance under different value of $U$. Specifically, we fix the AP antenna array dimension at $8 \times 8$ and set the number of active users to 30, while varying $U$ from 100 to 3000. Especially, we use `HP', `SP' and `EP' to denote hybrid preamble, superimposed preamble and embedded preamble schemes, respectively. The average detection error number (DEN) instead of DER is used to characterize the AUD performance for fairness. Under the GAMP-PCSBL-La algorithm, the performance trends of the three schemes are illustrated in Fig. \ref{U}. It can be observed that when $U$ is relatively small, HP, SP, and EP all achieve satisfactory detection and estimation performance. However, as $U$ increases, the performance of EP deteriorates significantly, as the growth rate of the sensing matrix dimension in EP is much higher than in the other two schemes. Initially, HP outperforms SP in terms of active user detection, but as $U$ increases, the detection performance of both schemes begins to converge. This is because, in high-$U$ scenarios, the detection performance of HP is primarily constrained by the rough AUD stage. Nevertheless, HP consistently outperforms SP in channel estimation by at least 5dB, and this performance gap increases with $U$. Overall, HP exhibits the best performance in massive UE scenarios.

\begin{figure*}[hb]
  \vspace{-1.2em}
	\centering
	\subfigure[]{\includegraphics[width=3.3in]{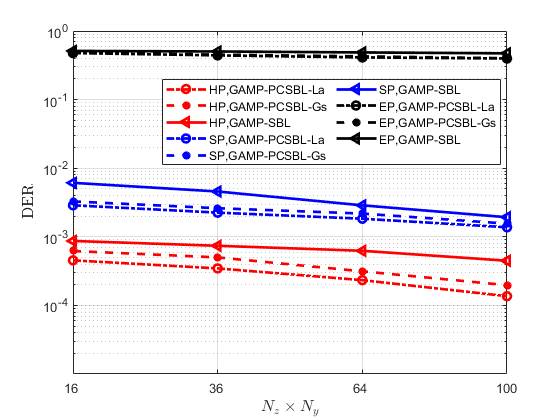}}
  \hspace{9mm}
  \subfigure[]{\includegraphics[width=3.3in]{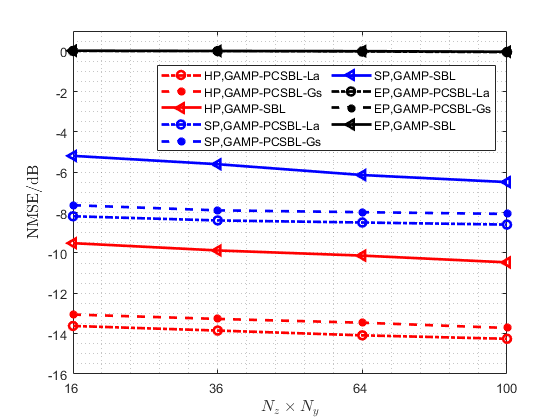}}
	\caption{Performance comparisons for massive random access schemes versus dimensions of antenna array: (a) DER; (b) NMSE.}
	\label{C}
\end{figure*}
\begin{figure*}[hb]
  \vspace{-0.8em}
	\centering
	\subfigure[]{\includegraphics[width=3.3in]{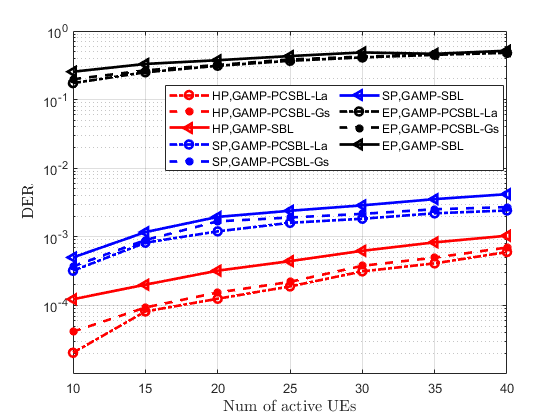}}
  \hspace{9mm}
  \subfigure[]{\includegraphics[width=3.3in]{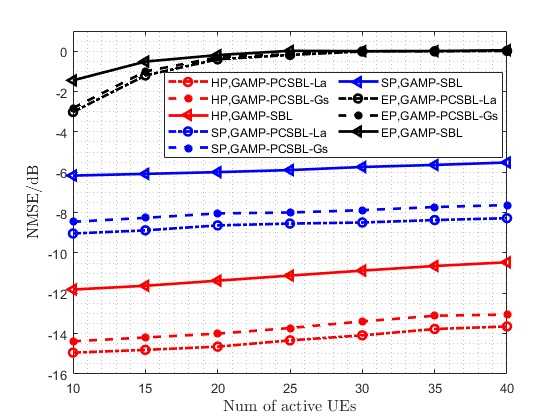}}
	\caption{Performance comparisons for massive random access schemes versus number of active UEs: (a) DER; (b) NMSE.}
	\label{E}
  \vspace{-1.9em}
\end{figure*}

In Fig. \ref{C}, with the number of active UEs fixed at 30, we simulated the impact of antenna array dimensions on the performance of massive random access schemes. Additionally, to avoid the complex computation of large matrix inversions, we limited our comparison to low-complexity GAMP-based algorithms to evaluate their performance in massive random access, thereby verifying the superiority of the proposed scheme. The results indicate that as the number of antennas increases, both the hybrid preamble scheme and the superimposed preamble scheme exhibit improved AUD and CE performance. Due to the absence of a rough activity detection step to reduce the dimensionality of the matrix to be estimated, the embedded preamble scheme alone, with its excessively large block-sparse channel matrix, fails to achieve effective AUD and CE results. It is evident that in massive random access, the proposed hybrid preamble scheme significantly outperforms the schemes that utilize either the superimposed or embedded preamble alone. Additionally, the simulation curves demonstrate that, compared to the GAMP-PCSBL-Gs and GAMP-SBL algorithms, the proposed GAMP-PCSBL-La algorithm more effectively captures the block-sparsity caused by fractional channel parameters, resulting in superior AUD and CE performance.

Similar conclusions can be drawn from the simulation curves in Fig. \ref{E}. With the antenna array dimension fixed at $8 \times 8$, we vary the number of active UEs from 10 to 40. The simulation results show that as the number of active UEs increases, the performance of all the massive random access schemes declines. This decline is attributed to the fact that more active UEs correspond to more non-zero elements, making the matrix less sparse. The proposed hybrid preamble scheme achieves significantly better AUD and CE performance compared to the other two preamble schemes when addressing the demands of massive random access. The larger dimension of the channel matrix to be estimated causes the embedded preamble scheme to fail when used alone. Moreover, the simulation curves in Fig. \ref{E} demonstrate that the proposed GAMP-PCSBL-La algorithm outperforms other iterative algorithms in block-sparse matrix recovery.

\vspace{-1em}
\section{Conclusion}
This paper proposes a hybrid preamble scheme for massive machine-type random access in high-mobility scenarios within cell-free massive MIMO systems using OTFS modulation. This scheme employs a superimposed preamble for rough AUD, then performs accurate AUD and CE based on the rough detected UE set and embedded preamble. By leveraging the advantages of both preamble schemes, the proposed hybrid preamble scheme achieves more precise detection and estimation with reduced preamble overhead. Additionally, a GAMP-PCSBL-La algorithm is introduced to estimate the channel matrix, effectively capturing the block-sparse characteristics of the channel caused by fractional channel parameters, while maintaining low computational complexity. Simulation results demonstrate that the proposed hybrid preamble scheme better meets the requirements for massive random access in OTFS-modulating cell-free massive MIMO systems, and that the GAMP-PCSBL-La algorithm is particularly well-suited for this scheme.

{\appendices
\section{}
By substituting equations (\ref{Xtfnm}) and (\ref{tau}), (\ref{nu}) into equation (\ref{Ybnm2}), we obtain:
\begin{small}
  \begin{equation}\label{104}
    \begin{gathered}
      \mathbf{y}_{u,b}\left[ {n,m} \right] = \frac{1}{T}\sum\limits_i {} {h_{u,b,i}}{\mathbf{a}_{u,b,i}}\sum\limits_{m'} {} \frac{1}{{\sqrt {NM} }}\sum\limits_{k'} {} \sum\limits_{l'} {} {X_u^{DD}}\left[ {k',l'} \right]\hfill\\
       {e^{ - j2\pi \left( {\frac{{m'l'}}{M} - \frac{{nk'}}{N}} \right)}}{e^{ - j2\pi \frac{{m'{l_i}}}{M}}}
      {e^{ - j2\pi \frac{{{l_i}\left( {{k_i} + {{\tilde k}_i}} \right)}}{{NM}}}}{e^{j2\pi \frac{{n({k_i} + {{\tilde k}_i})}}{N}}} \hfill \\
      \sum\limits_{p = {l_{u,b,i}}}^M {} \frac{1}{{M\Delta f}}{e^{ - j2\pi \frac{p}{M}(m - m' - \frac{{{k_{u,b,i}} + {{\tilde k}_{u,b,i}}}}{N})}} +\hfill \\
       \frac{1}{T}\sum\limits_i {} {h_{u,b,i}}{\mathbf{a}_{u,b,i}}\sum\limits_{m'} {} \frac{1}{{\sqrt {NM} }}\sum\limits_{k'} {} 
      \sum\limits_{l'} {} {X^{DD}}\left[ {k',l'} \right]{e^{ - j2\pi \left( {\frac{{m'l'}}{M} - \frac{{nk'}}{N}} \right)}} \hfill \\
      {e^{ - j2\pi \frac{{m'{l_i}}}{M}}}{e^{ - j2\pi \frac{{{l_i}\left( {{k_{u,b,i}} + {{\tilde k}_{u,b,i}}} \right)}}{{NM}}}}{e^{j2\pi \frac{{\left( {n - 1} \right)({k_{u,b,i}} + {{\tilde k}_{u,b,i}})}}{N}}}\hfill \\
      \sum\limits_{p = 0}^{{l_{u,b,i}}-1} {} \frac{1}{{M\Delta f}}{e^{ - j2\pi \frac{p}{M}(m - m' - \frac{{{k_{u,b,i}} + {{\tilde k}_{u,b,i}}}}{N})}} + \mathbf{n}_{u,b}[n,m]. \hfill \\
    \end{gathered}
  \end{equation}
\end{small}
Furthermore, substituting equation (\ref{Yddkl}) into the above equation then we get:
\begin{small}
  \begin{align}\label{105}
      &{\mathbf{y}_{u,b}^{DD}}\left[ {k,l} \right] = \sum\limits_n {} \sum\limits_m {} \sum\limits_i {} {h_{u,b,i}}{\mathbf{a}_{u,b,i}}\sum\limits_{m'} {} \frac{1}{{NM}}\sum\limits_{k'} {} {\sum\limits_{l'} X_u^{DD}}\left[ {k',l'} \right] \notag \\
      &{e^{ - j2\pi \left( {\frac{{m'l'}}{M} - \frac{{nk'}}{N}} \right)}}{e^{ - j2\pi \frac{{m'{l_{u,b,i}}}}{M}}}{e^{ - j2\pi \frac{{{l_{u,b,i}}\left( {{k_{u,b,i}} + {{\tilde k}_{u,b,i}}} \right)}}{{NM}}}} \notag \\
      &{e^{j2\pi \frac{{n({k_{u,b,i}} + {{\tilde k}_{u,b,i}})}}{N}}}\sum\limits_{p = {l_{u,b,i}}}^M {}\frac{1}{M}{e^{ - j2\pi \frac{p}{M}(m - m' - \frac{{{k_{u,b,i}} + {{\tilde k}_{u,b,i}}}}{N})}} \notag \\
      &{e^{j2\pi \left( {\frac{{ml}}{M} - \frac{{nk}}{N}} \right)}} + \sum\limits_n {} \sum\limits_m {} \sum\limits_i {} {h_{u,b,i}}{\mathbf{a}_{u,b,i}}\sum\limits_{m'} \frac{1}{{NM}}{e^{j2\pi \left( {\frac{{ml}}{M} - \frac{{nk}}{N}} \right)}}\sum\limits_{k'} {} \notag \\
      &{\sum\limits_{l'} X ^{DD}}\left[ {k',l'} \right]{e^{ - j2\pi \left( {\frac{{m'l'}}{M} - \frac{{\left( {n - 1} \right)k'}}{N}} \right)}}   {e^{ - j2\pi \frac{{{l_{u,b,i}}\left( {{k_{u,b,i}} + {{\tilde k}_{u,b,i}}} \right)}}{{NM}}}}\notag \\
      &{e^{ - j2\pi \frac{{m'{l_{u,b,i}}}}{M}}}{e^{j2\pi \frac{{n({k_{u,b,i}} + {{\tilde k}_{u,b,i}})}}{N}}}\sum\limits_{p = 0}^{{l_{u,b,i}}} {} \frac{1}{M}{e^{ - j2\pi \frac{p}{M}(m - m' - \frac{{{k_{u,b,i}} + {{\tilde k}_{u,b,i}}}}{N})}}  \notag \\
       &= \sum\limits_i {} {h_{u,b,i}}{\mathbf{a}_{u,b,i}}\sum\limits_{k'} {} {\sum\limits_{l'} X ^{DD}}\left[ {k',l'} \right]{e^{ - j2\pi \frac{{{l_i}\left( {{k_{u,b,i}} + {{\tilde k}_{u,b,i}}} \right)}}{{NM}}}} \notag \\
       &\sum\limits_{p = {l_{u,b,i}}}^M {} {e^{j2\pi \frac{p}{M}(\frac{{{k_{u,b,i}} + {{\tilde k}_{u,b,i}}}}{N})}} 
      \sum\limits_{m'} {} \frac{1}{M}{e^{ - j2\pi \frac{{m'}}{M}(l' + {l_{u,b,i}} - p)}} \notag \\
      &\sum\limits_m {} \frac{1}{M}{e^{ - j2\pi \frac{m}{M}(p - l)}}\sum\limits_n {}\frac{1}{N}{e^{ - j2\pi \frac{n}{N}(k - {k_{u,b,i}} - {{\tilde k}_{u,b,i}} - k')}} + \notag \\
      &\sum\limits_i {} {h_{u,b,i}}{\mathbf{a}_{u,b,i}}\sum\limits_{k'} {} {\sum\limits_{l'} X ^{DD}}\left[ {k',l'} \right]{e^{ - j2\pi \frac{{k'}}{N}}}\notag \\
      &\sum\limits_{p = 0}^{{l_{u,b,i}}-1} {} {e^{j2\pi \frac{{p - {l_{u,b,i}}}}{M}(\frac{{{k_{u,b,i}} + {{\tilde k}_{u,b,i}}}}{N})}}\sum\limits_m {} \frac{1}{M}{e^{ - j2\pi \frac{m}{M}(p - l)}} \notag \\
      &\sum\limits_{m'} {} \frac{1}{M}{e^{ - j2\pi \frac{{m'}}{M}(l' + {l_{u,b,i}} - p)}}\sum\limits_n {} \frac{1}{N}{e^{ - j2\pi \frac{n}{N}(k - {k_{u,b,i}} - {{\tilde k}_{u,b,i}} - k')}}.
    \end{align}
\end{small}
We define ${\bm{\Pi }}_N\left( {x - a} \right) = \frac{1}{N}\sum\limits_{n = 0}^{N - 1} {} {e^{ - j2\pi \frac{n}{N}\left( {x - a} \right)}}$, thus, the above equation can be expressed as:
\begin{small}
  \begin{equation}\label{106}
    \begin{gathered}
      {\mathbf{y}_{b,b}^{DD}}\left[ {k,l} \right] = \sum\limits_i {} {h_{u,b,i}}{\mathbf{a}_{u,b,i}}\sum\limits_{k'} {} {\sum\limits_{l'} X_u^{DD}}\left[ {k',l'} \right] \hfill \\
      \sum\limits_{p = {l_{u,b,i}}}^M {} {e^{j2\pi \frac{{\left( {p - {l_{u,b,i}}} \right)\left( {{k_{u,b,i}} + {{\tilde k}_{u,b,i}}} \right)}}{{MN}}}}\delta \left( {l' + {l_{u,b,i}} - p} \right) \hfill \\
      \delta \left( {p - l} \right){{\bm{\Pi }}_N}\left( {k - {k_{u,b,i}} - {{\tilde k}_{u,b,i}} - k'} \right) + \sum\limits_i {} {h_{u,b,i}}{\mathbf{a}_{u,b,i}}\sum\limits_{k'} {} \sum\limits_{l'} \hfill \\
      X_u^{DD}\left[ {k',l'} \right]{e^{ - j2\pi \frac{{k'}}{N}}} 
      \sum\limits_{p = 0}^{{l_{u,b,i}} - 1} {} {e^{j2\pi \frac{{\left( {p - {l_{u,b,i}}} \right)\left( {{k_{u,b,i}} + {{\tilde k}_{u,b,i}}} \right)}}{{MN}}}}\hfill \\
      \delta \left( {l' + {l_{u,b,i}} - p} \right)\delta \left( {p - l} \right){{\bm{\Pi }}_N}\left( {k - {k_{u,b,i}} - {{\tilde k}_{u,b,i}} - k'} \right). \hfill \\
    \end{gathered}
  \end{equation}
\end{small}
Analyzing the equation (\ref{106}), and using the properties of the Dirac delta function, ${\mathbf{y}_{u,b}}^{DD}\left[ {k,l} \right]$ can be written as a segment function. When ${l_{u,b,i}} \le l$, we have:
\begin{small}
\begin{equation}\label{107}
\begin{gathered}
  {\mathbf{y}_{u,b}}^{DD}\left[ {k,l} \right] = \sum_i{} {h_{u,b,i}}{\mathbf{a}_{u,b,i}}\sum\limits_{k'} {} {X_u^{DD}}\left[ {k',l - {l_{u,b,i}}} \right]\hfill \\
  {e^{j2\pi \frac{{\left( {l - {l_{u,b,i}}} \right)\left( {{k_{u,b,i}} + {{\tilde k}_{u,b,i}}} \right)}}{{NM}}}} {{\bm{\Pi }}_N}\left( {k - {k_{u,b,i}} - {{\tilde k}_{u,b,i}} - k'} \right) \hfill \\
  \mathop  \approx \limits^a \sum_i{h_{u,b,i}}{\mathbf{a}_{u,b,i}}\sum\limits_{k''} {} {X_u^{DD}}\left[ {k - k'',l - {l_{u,b,i}}} \right] \hfill \\
  {e^{j2\pi \frac{{\left( {l - {l_{u,b,i}}} \right)\left( {{k_{u,b,i}} + {{\tilde k}_{u,b,i}}} \right)}}{{NM}}}}\frac{1}{N}\frac{{1 - {e^{j2\pi {{\tilde k}_{u,b,i}}}}}}{{1 - {e^{ - j2\pi \frac{{k'' - {k_{u,b,i}} - {{\tilde k}_{u,b,i}}}}{N}}}}} \hfill \\
\end{gathered}
\end{equation}
\end{small}
Here, the approximate equality $a$ retains only the $2\varepsilon + 1$ integer points near the extremum to approximate ${{\bm{\Pi }}_N}\left( {k - {k_{u,b,i}} - {{\tilde k}_{u,b,i}} - k'} \right)$, where $k'' \in \left[ {{k_i} - \varepsilon ,{k_i} + \varepsilon } \right]$. Similarly, when ${l_{u,b,i}} > l$, we have
\begin{small}
\begin{align}\label{108}
  &{{\mathbf{y}_{u,b}}^{DD}}\left[ {k,l} \right] =  \sum_i{} {h_{u,b,i}}{\mathbf{a}_{u,b,i}}\sum\limits_{k'} {} {X_u^{DD}}\left[ {k',l - {l_{u,b,i}}} \right]{e^{ - j2\pi \frac{{k'}}{N}}} \notag \\
  &{e^{j2\pi \frac{{\left( {l - {l_{u,b,i}}} \right)\left( {{k_{u,b,i}} + {{\tilde k}_{u,b,i}}} \right)}}{{NM}}}} 
  {{\bm{\Pi }}_N}\left( {k - {k_{u,b,i}} - {{\tilde k}_{u,b,i}} - k'} \right)\notag \\
  &\approx  \sum_i{h_{u,b,i}}{\mathbf{a}_{u,b,i}}\sum\limits_{k''} {} {X^{DD}}\left[ {k - k'',l - {l_{u,b,i}}} \right]{e^{ - j2\pi \frac{{k - k''}}{N}}} \notag \\
  &{e^{j2\pi \frac{{\left( {l - {l_{u,b,i}}} \right)\left( {{k_{u,b,i}} + {{\tilde k}_{u,b,i}}} \right)}}{{NM}}}}\frac{1}{N}\frac{{1 - {e^{j2\pi {{\tilde k}_{u,b,i}}}}}}{{1 - {e^{ - j2\pi \frac{{k'' - {k_{u,b,i}} - {{\tilde k}_{u,b,i}}}}{N}}}}}.
\end{align}
\end{small}
Combining the above derivations, we obtain equation (\ref{Ydd}).

\section{}
In the case where $M$ is very small, the received signal in the time-frequency domain can be expressed as:
\begin{small}
\begin{equation}\label{109}
\begin{gathered}
  \mathbf{y}_{u,b}\left[ {n,m} \right] = \frac{1}{T}\sum\limits_i {} {h_{u,b,i}}{\mathbf{a}_{u,b,i}}\sum\limits_{m'} {} \frac{1}{{\sqrt {NM} }}\sum\limits_{k'} {} \sum\limits_{l'} {} {X_u^{DD}}\left[ {k',l'} \right] \hfill \\
  {e^{ - j2\pi \left( {\frac{{m'l'}}{M} - \frac{{nk'}}{N}} \right)}} {e^{ - j2\pi \frac{{{{\tilde l}_i}\left( {{k_{u,b,i}} + {{\tilde k}_{u,b,i}}} \right)}}{{NM}}}}{e^{ - j2\pi \frac{{m'{{\tilde l}_{u,b,i}}}}{M}}}{e^{j2\pi \frac{{n({k_{u,b,i}} + {{\tilde k}_i})}}{N}}}  \hfill \\
  \left( \sum\limits_{p = 0}^M {} \frac{1}{{M\Delta f}}{e^{ - j2\pi \frac{p}{M}(m - m' - \frac{{{k_{u,b,i}} + {{\tilde k}_{u,b,i}}}}{N})}} -\right.\hfill \\
  \left. \int_0^{{\tau _{u,b,i}}} {} {e^{ - j2\pi \Delta ft(m - m' - \frac{{{\nu _{u,b,i}}}}{{\Delta f}})}}dt \right) \hfill \\
  \mathop  \approx \limits^a \frac{1}{T}\sum\limits_i {} {h_{u,b,i}}{\mathbf{a}_{u,b,i}}\sum\limits_{m'} {} \frac{1}{{\sqrt {NM} }}\sum\limits_{k'} {} \sum\limits_{l'} {} {X_u^{DD}}\left[ {k',l'} \right]{e^{ - j2\pi \frac{{m'{{\tilde l}_{u,b,i}}}}{M}}}\hfill \\
  {e^{ - j2\pi \left( {\frac{{m'l'}}{M} - \frac{{nk'}}{N}} \right)}}{e^{ - j2\pi \frac{{{{\tilde l}_{u,b,i}}\left( {{k_{u,b,i}} + {{\tilde k}_{u,b,i}}} \right)}}{{NM}}}}{e^{j2\pi \frac{{n({k_{u,b,i}} + {{\tilde k}_{u,b,i}})}}{N}}} \hfill \\
  \sum\limits_{p = 0}^M {} \frac{1}{{M\Delta f}}{e^{ - j2\pi \frac{p}{M}(m - m' - \frac{{{k_{u,b,i}} + {{\tilde k}_{u,b,i}}}}{N})}}. \hfill \\
\end{gathered}
\end{equation}
\end{small}
In our system, assuming that the delay parameter is much smaller than the duration of one symbol, we can roughly establish the approximation $a$ in the above equation. Similar to the derivation in Appendix A, we can substitute equations (\ref{Xtfnm}), (\ref{tau}), (\ref{nu}), and (\ref{Yddkl}) to obtain:
\begin{small}
\begin{align*}
  &{\mathbf{y}_{u,b}^{DD}}\left[ {k,l} \right] \approx \sum\limits_n {} \sum\limits_m {} \sum\limits_i {} {h_{u,b,i}}{\mathbf{a}_{u,b,i}}\sum\limits_{m'} {} \frac{1}{{NM}}\sum\limits_{k'} {} {\sum\limits_{l'} X_u ^{DD}}\left[ {k',l'} \right]\\
  &{e^{ - j2\pi \left( {\frac{{m'l'}}{M} - \frac{{nk'}}{N}} \right)}}{e^{ - j2\pi \frac{{m'{{\tilde l}_i}}}{M}}}{e^{ - j2\pi \frac{{{{\tilde l}_i}\left( {{k_{u,b,i}} + {{\tilde k}_{u,b,i}}} \right)}}{{NM}}}}{e^{j2\pi \frac{{n({k_{u,b,i}} + {{\tilde k}_{u,b,i}})}}{N}}} \\
  &\sum\limits_{p = {l_{u,b,i}}}^M {} \frac{1}{M}{e^{j2\pi \left( {\frac{{ml}}{M} - \frac{{nk}}{N}} \right)}}{e^{ - j2\pi \frac{p}{M}(m - m' - \frac{{{k_{u,b,i}} + {{\tilde k}_{u,b,i}}}}{N})}} \\
  & = \sum\limits_i {} {h_{u,b,i}}{\mathbf{a}_{u,b,i}}\sum\limits_{k'} {} {\sum\limits_{l'} X ^{DD}}\left[ {k',l'} \right]{e^{ - j2\pi \frac{{{{\tilde l}_{u,b,i}}\left( {{k_{u,b,i}} + {{\tilde k}_{u,b,i}}} \right)}}{{NM}}}}\\
  &\sum\limits_{p = {l_{u,b,i}}}^M {} {e^{j2\pi \frac{p}{M}(\frac{{{k_{u,b,i}} + {{\tilde k}_{u,b,i}}}}{N})}} 
  \sum\limits_{m'} {} \frac{1}{M}{e^{ - j2\pi \frac{{m'}}{M}(l' + {{\tilde l}_{u,b,i}} - p)}} \\
  &\sum\limits_m {} \frac{1}{M}{e^{ - j2\pi \frac{m}{M}(p - l)}}\sum\limits_n {} \frac{1}{N}{e^{ - j2\pi \frac{n}{N}(k - {k_{u,b,i}} - {{\tilde k}_{u,b,i}} - k')}}\\
  & = \sum\limits_i {}{h_{u,b,i}}{\mathbf{a}_{u,b,i}}\sum\limits_{k'} {} {\sum\limits_{l'} X ^{DD}}\left[ {k',l'} \right]{e^{ - j2\pi \frac{{{{\tilde l}_{u,b,i}}\left( {{k_{u,b,i}} + {{\tilde k}_{u,b,i}}} \right)}}{{NM}}}}  \\
  &\sum\limits_{p = {l_{u,b,i}}}^M {} {e^{j2\pi \frac{p}{M}(\frac{{{k_{u,b,i}} + {{\tilde k}_{u,b,i}}}}{N})}}{{\bm{\Pi }}_M}\left( {l' + {{\tilde l}_{u,b,i}} - p} \right)\delta \left( {p - l} \right) \\
  &{{\bm{\Pi }}_N}\left( {k - {k_{u,b,i}} - {{\tilde k}_{u,b,i}} - k'} \right)\\
  & = \sum\limits_i {} {h_{u,b,i}}{\mathbf{a}_{u,b,i}}\sum\limits_{k'} {} {\sum\limits_{l'} X ^{DD}}\left[ {k',l'} \right]{e^{j2\pi \frac{{(l - {{\tilde l}_{u,b,i}})\left( {{k_{u,b,i}} + {{\tilde k}_{u,b,i}}} \right)}}{{NM}}}}  \\
  &{{\bm{\Pi }}_M}\left( {l' + {{\tilde l}_{u,b,i}} - l} \right){{\bm{\Pi }}_N}\left( {k - {k_{u,b,i}} - {{\tilde k}_{u,b,i}} - k'} \right) \\
  & \approx \frac{1}{{NM}}\sum\limits_i {} {h_{u,b,i}}{\mathbf{a}_{u,b,i}}\sum\limits_{k''} {} {X^{DD}}\left[ {k - k'',l} \right]\frac{{1 - {e^{ - j2\pi {{\tilde l}_{u,b,i}}}}}}{{1 - {e^{ - j2\pi \frac{{{{\tilde l}_{u,b,i}}}}{M}}}}} \\
  &{e^{j2\pi \frac{{\left( {l - {{\tilde l}_{u,b,i}}} \right)\left( {{k_{u,b,i}} + {{\tilde k}_{u,b,i}}} \right)}}{{NM}}}}\frac{{1 - {e^{j2\pi {{\tilde k}_{u,b,i}}}}}}{{1 - {e^{ - j2\pi \frac{{k'' - {k_{u,b,i}} - {{\tilde k}_{u,b,i}}}}{N}}}}}.
\end{align*}
\end{small}
Then we can derive equation (\ref{Ydd1}) based on above results.

\section{}
For the case where $M$ is small, the derivation of equation (\ref{Yubp1}) can be found in Appendix B. For the case where $M$ is large, according to equation (\ref{Ybnm2}), we have:
\begin{small}
\begin{align*}
  &{\mathbf{y}_{u,b}}\left[ {\frac{{n'}}{\alpha },m'} \right] = \frac{1}{T}\sum\limits_i {}{h_{u,b,i}}{\mathbf{a}_{u,b,i}}\sum\limits_{m''} {} {{X}}_{u,1}^{TF}\left[\frac{{n'}}{\alpha },m''\right]  {e^{ - j2\pi m''\Delta f{\tau _{u,b,i}}}}  \\
  &{e^{ - j2\pi {\nu _{u,b,i}}{\tau _{u,b,i}}}}{e^{j2\pi {\nu _{u,b,i}}\frac{{n'}}{\alpha }T}}\int_{{\tau _{u,b,i}}}^T {} {e^{ - j2\pi \Delta ft(m' - m'' - \frac{{{\nu _{u,b,i}}}}{{\Delta f}})}}dt   \\
  &+\frac{1}{T}\sum\limits_i {} {h_{u,b,i}}{\mathbf{a}_{u,b,i}}\sum\limits_{m''} {} {{X}}_{u,1}^{TF}\left[\frac{{n'}}{\alpha } - 1,m''\right]{e^{ - j2\pi m''\Delta f{\tau _{u,b,i}}}}  \\
  &{e^{ - j2\pi m''\Delta fT}}{e^{ - j2\pi {\nu _{u,b,i}}{\tau _{u,b,i}}}}{e^{j2\pi {\nu _{u,b,i}}\frac{{n'}}{\alpha }T}} \\
  &\int_0^{{\tau _{u,b,i}}} {} {e^{ - j2\pi \Delta ft(m' - m'' - \frac{{{\nu _{u,b,i}}}}{{\Delta f}})}}dt  \\
  &\mathop  = \limits^a \frac{1}{T}\sum\limits_i {} {h_{u,b,i}}{\mathbf{a}_{u,b,i}}\sum\limits_{m''} {} {{X}}_{u,1}^{TF}\left[\frac{{n'}}{\alpha },m''\right]{e^{ - j2\pi m''\Delta f{\tau _{u,b,i}}}}\\
  &{e^{ - j2\pi {\nu _{u,b,i}}{\tau _{u,b,i}}}}{e^{j2\pi {\nu _{u,b,i}}\frac{{n'}}{\alpha }T}} \sum\limits_{p = 0}^{M - 1} {\frac{1}{{M\Delta f}}{e^{ - j2\pi \frac{p}{M}(m' - m'' - \frac{{{\nu _{u,b,i}}}}{{\Delta f}})}}}  \\
  &+ \frac{1}{T}\sum\limits_i {} {h_{u,b,i}}{\mathbf{a}_{u,b,i}}\sum\limits_{m''} {} \sum\limits_{p = 0}^{{l_{u,b,i}}} {\frac{1}{{M\Delta f}}{e^{ - j2\pi \frac{p}{M}(m' - m'' - \frac{{{\nu _{u,b,i}}}}{{\Delta f}})}}}   \\
  &{e^{ - j2\pi m''\Delta f{\tau _{u,b,i}}}}{e^{ - j2\pi {\nu _{u,b,i}}{\tau _{u,b,i}}}}{e^{j2\pi {\nu _{u,b,i}}\frac{{n'}}{\alpha }T}}  \left( {{X}}_{u,1}^{TF}\left[\frac{{n'}}{\alpha } - 1,m''\right]\right.\\
  &\left. {e^{ - j2\pi m''\Delta fT}} - {{X}}_{u,1}^{TF}\left[\frac{{n'}}{\alpha },m''\right] \right).
\end{align*}
\end{small}
Assuming the time-frequency domain symbols ${\mathbf{X}}_{u,1}^{TF}[n',m']$ follow a zero-mean Gaussian distribution, according to the central limit theorem, the ratio of variances between the first and second terms on the right-hand side of the equation $a$ is $\frac{M}{{2\bar l}}$, where $\bar l$ is the expected value of the delay quantization value ${l_{u,b,i}}$. Typically, delays are assumed to be uniformly randomly distributed, so $2\bar l = {l_{\max }}$ and ${l_{\max }} \ll M$ It can be considered that the first term on the right-hand side of equation $a$ dominates the numerical value. By placing the second term of equation $a$ into the noise, we obtain the equation (\ref{Yubp1}).

\section{}
According to equations (\ref{78}) and (\ref{79}), the posterior mean of ${x_{i,j}}$ can be expressed as:
\begin{small}
\begin{equation}\label{110}
\begin{gathered}
  {{\hat x}_{i,j}}(t + 1) = \int {{x_{i,j}}p\left( {{x_{i,j}}|{\mathbf{Y}},{\mathbf{\tau }},\gamma } \right)d} {x_{i,j}} \hfill \\
   = \frac{{\sqrt {2\pi u_{i,j}^r(t)} }}{{{\psi _{i,j}}(t)}}\left[ {e^{ - \xi _{i,j}^ + \left( t \right)}}{\psi _1}\left( {\varphi _{i,j}^ + \left( t \right),u_{i,j}^r(t)} \right) - {e^{ - \xi _{i,j}^ - \left( t \right)}} \right. \hfill \\
  \left. {\psi _1}\left( { - \varphi _{i,j}^ - \left( t \right),u_{i,j}^r(t)} \right) \right], \hfill \\
\end{gathered}
\end{equation}
\end{small}
where
\begin{small}
\begin{equation}\label{111}
\begin{gathered}
  {\psi _1}\left( {\varphi ,u} \right) = \frac{1}{{\sqrt {2\pi u} }}\int_0^{ + \infty } {} t\exp \left\{ { - \frac{{{{\left( {t - \varphi } \right)}^2}}}{{2u}}} \right\}dt \hfill \\
   = \varphi Q\left( { - \frac{\varphi }{{\sqrt u }}} \right) + \frac{u}{{\sqrt {2\pi u} }}\exp \left\{ { - \frac{{{\varphi ^2}}}{{2u}}} \right\}. \hfill \\
\end{gathered}
\end{equation}
\end{small}
Then we have
\begin{small}
\begin{align}\label{112}
  &{\hat x_{i,j}}(t + 1) = \frac{{\sqrt {2\pi u_{i,j}^r(t)} }}{{{\psi _{i,j}}(t)}}\left[ {e^{ - \xi _{i,j}^ + \left( t \right)}}\varphi _{i,j}^ + \left( t \right)Q\left( {{{ - \varphi _{i,j}^ + \left( t \right)}}/{{\sqrt {u_{i,j}^r(t)} }}} \right) \right. \notag \\
   &+ {e^{ - \xi _{i,j}^ - \left( t \right)}}\varphi _{i,j}^ - \left( t \right)Q\left( {{{\varphi _{i,j}^ - \left( t \right)}}/{{\sqrt {u_{i,j}^r(t)} }}} \right) \notag \\
   &\left.+ \frac{{u_{i,j}^r(t)}}{{\sqrt {2\pi u_{i,j}^r(t)} }}\left( {{e^{ - \xi _{i,j}^ + \left( t \right) - \frac{{{{\left( {\varphi _{i,j}^ + \left( t \right)} \right)}^2}}}{{2u_{i,j}^r(t)}}}} - {e^{ - \xi _{i,j}^ - \left( t \right) - \frac{{{{\left( {\varphi _{i,j}^ - \left( t \right)} \right)}^2}}}{{2u_{i,j}^r(t)}}}}} \right) \right]. 
\end{align}
\end{small}
From equations (\ref{82}) to (\ref{85}), we can get that
\begin{small}
\begin{equation}\label{113}
\xi _{i,j}^ + \left( t \right) + \frac{{{{\left( {\varphi _{i,j}^ + \left( t \right)} \right)}^2}}}{{2u_{i,j}^r(t)}} = \xi _{i,j}^ - \left( t \right) + \frac{{{{\left( {\varphi _{i,j}^ - \left( t \right)} \right)}^2}}}{{2u_{i,j}^r(t)}} = \frac{{{{\left( {{{\hat r}_{i,j}}(t)} \right)}^2}}}{{2u_{i,j}^r(t)}}.
\end{equation}
\end{small}
The last two terms of equation (\ref{112}) can be eliminated, resulting in equation (\ref{hatxij}). We define
\begin{small}
\begin{equation}\label{114}
\begin{gathered}
  {\chi _{i,j}}(t + 1) = \int {} x_{i,j}^2p\left( {{x_{i,j}}|{\mathbf{Y}},{\mathbf{\tau }},\gamma } \right)d{x_{i,j}} \hfill \\
   = \frac{{\sqrt {2\pi u_{i,j}^r(t)} }}{{{\psi _{i,j}}(t)}}\left[ {e^{ - \xi _{i,j}^ + \left( t \right)}}{\psi _2}\left( {\varphi _{i,j}^ + \left( t \right),u_{i,j}^r(t)} \right) + {e^{ - \xi _{i,j}^ - \left( t \right)}} \right.\hfill \\
  \left. {\psi _2}\left( { - \varphi _{i,j}^ - \left( t \right),u_{i,j}^r(t)} \right) \right], \hfill \\
\end{gathered}
\end{equation}
\end{small}
where
\begin{small}
\begin{equation}\label{115}
{\psi _2}\left( {\varphi ,u} \right) = \frac{1}{{\sqrt {2\pi u} }}\int_0^{ + \infty } {} {t^2}\exp \left\{ { - \frac{{{{\left( {t - \varphi } \right)}^2}}}{{2u}}} \right\}dt.
\end{equation}
\end{small}
First we have
\begin{small}
\begin{equation}\label{116}
\begin{array}{*{20}{c}}
  {g\left( t \right) = \exp \left\{ { - \frac{{{{\left( {t - \varphi } \right)}^2}}}{{2u}}} \right\}}& \to &{g'\left( t \right) =  - \frac{{t - \varphi }}{u}g\left( t \right)} \\
  {f\left( t \right) = t}& \to &{f'\left( t \right) = 1}
\end{array},
\end{equation}
\end{small}
using the fact that
\begin{small}
\begin{equation}\label{117}
\int_0^{ + \infty } {f\left( t \right)g'\left( t \right)dt}  = \left. {f\left( t \right)g\left( t \right)} \right|_0^{ + \infty } - \int_0^{ + \infty } {f'\left( t \right)g\left( t \right)dt},
\end{equation}
\end{small}
and $\left. {f\left( t \right)g\left( t \right)} \right|_0^{ + \infty } = 0$ to get
\begin{small}
\begin{equation}\label{118}
\int_0^{ + \infty } {} \frac{{t\left( {t - \varphi } \right)}}{u}\exp \left\{ { - \frac{{{{\left( {t - \varphi } \right)}^2}}}{{2u}}} \right\}dt = \int_0^{ + \infty } {} \exp \left\{ { - \frac{{{{\left( {t - \varphi } \right)}^2}}}{{2u}}} \right\}dt.
\end{equation}
\end{small}
In the right-hand side of equation (\ref{118}), we set $x = \left( {t - \varphi } \right)/\sqrt u $ and substitute the definitions of ${\psi _1}\left( {\varphi ,u} \right)$ and ${\psi _2}\left( {\varphi ,u} \right)$ into the left-hand side, yielding:
\begin{small}
\begin{equation}\label{119}
\frac{{\sqrt {2\pi u} }}{u}{\psi _2}\left( {\varphi ,u} \right) - \frac{{\varphi \sqrt {2\pi u} }}{u}{\psi _1}\left( {\varphi ,u} \right) = \sqrt {2\pi u} Q\left( {\frac{{ - \gamma }}{{\sqrt u }}} \right).
\end{equation}
\end{small}
Then we get
\begin{small}
\begin{equation}\label{120}
{\psi _2}\left( {\varphi ,u} \right) = \varphi {\psi _1}\left( {\varphi ,u} \right) + uQ\left( {\frac{{ - \gamma }}{{\sqrt u }}} \right).
\end{equation}
\end{small}
With the definition of ${\psi _1}\left( {\varphi ,u} \right)$, it can be obtained that
\begin{small}
\begin{equation}\label{121}
\begin{gathered}
  {e^{ - \xi _{i,j}^ + \left( t \right)}}{\psi _2}\left( {\varphi _{i,j}^ + \left( t \right),u_{i,j}^r(t)} \right) =  \hfill \\
  \left( {{{\left( {\varphi _{i,j}^ + \left( t \right)} \right)}^2} + u_{i,j}^r(t)} \right){e^{ - \xi _{i,j}^ + \left( t \right)}}Q\left( {\frac{{ - \varphi _{i,j}^ + \left( t \right)}}{{\sqrt {u_{i,j}^r(t)} }}} \right) \hfill \\
   + \frac{{u_{i,j}^r(t)\varphi _{i,j}^ + \left( t \right)}}{{\sqrt {2\pi u_{i,j}^r(t)} }}\exp \left\{ { - \frac{{{{\left( {{{\hat r}_{i,j}}(t)} \right)}^2}}}{{2u_{i,j}^r(t)}}} \right\}, \hfill \\
\end{gathered}
\end{equation}
\end{small}
\begin{small}
\begin{equation}\label{122}
\begin{gathered}
  {e^{ - \xi _{i,j}^ - \left( t \right)}}{\psi _2}\left( { - \varphi _{i,j}^ - \left( t \right),u_{i,j}^r(t)} \right) =  \hfill \\
  \left( {{{\left( {\varphi _{i,j}^ - \left( t \right)} \right)}^2} + u_{i,j}^r(t)} \right){e^{ - \xi _{i,j}^ - \left( t \right)}}Q\left( {\frac{{ - \varphi _{i,j}^ - \left( t \right)}}{{\sqrt {u_{i,j}^r(t)} }}} \right) \hfill \\
   + \frac{{u_{i,j}^r(t)\varphi _{i,j}^ - \left( t \right)}}{{\sqrt {2\pi u_{i,j}^r(t)} }}\exp \left\{ { - \frac{{{{\left( {{{\hat r}_{i,j}}(t)} \right)}^2}}}{{2u_{i,j}^r(t)}}} \right\}. \hfill \\
\end{gathered}
\end{equation}
\end{small}
Combining equations (\ref{84}) and (\ref{85}), and substituting (\ref{121}) and (\ref{122}) into (\ref{114}), finally using the variance definition $u_{i,j}^x(t + 1) = {\chi _{i,j}}(t + 1) - {\left( {{{\hat x}_{i,j}}(t + 1)} \right)^2}$, we obtain the result of equation (\ref{uijx}).
}

\bibliographystyle{ieeetr}
\bibliography{refpaper}

\vfill

\end{document}